\documentclass[12pt,a4paper,superscriptaddress,floatfix]{article}

\usepackage{graphicx,pdflscape}
\usepackage{listings}

\usepackage{multirow}
\usepackage{parskip}
\newtheorem{thm}{Theorem}

\usepackage[hypcap]{caption}
\usepackage[paper=letterpaper,margin=1in]{geometry}
\usepackage{lscape}
\usepackage{amsmath,amssymb,amsfonts,epsfig,cite,setspace,bigstrut,longtable,array,breqn}

\pdfoutput=1

\begin{document}

\vskip 0.25in

\newcommand{\todo}[1]{{\bf ?????!!!! #1
?????!!!!}\marginpar{$\Longleftarrow$}}
\newcommand{\fref}[1]{Figure~\ref{#1}}
\newcommand{\tref}[1]{Table~\ref{#1}}
\newcommand{\sref}[1]{\S~\ref{#1}}
\newcommand{\nn}{\nonumber}
\newcommand{\tr}{\mathop{\rm Tr}}
\newcommand{\comment}[1]{}

\newcommand{\cM}{{\cal M}}
\newcommand{\cW}{{\cal W}}
\newcommand{\cN}{{\cal N}}
\newcommand{\cH}{{\cal H}}
\newcommand{\cK}{{\cal K}}
\newcommand{\cZ}{{\cal Z}}
\newcommand{\cO}{{\cal O}}
\newcommand{\cB}{{\cal B}}
\newcommand{\cC}{{\cal C}}
\newcommand{\cD}{{\cal D}}
\newcommand{\cE}{{\cal E}}
\newcommand{\cF}{{\cal F}}
\newcommand{\cX}{{\cal X}}
\newcommand{\IA}{\mathbb{A}}
\newcommand{\IP}{\mathbb{P}}
\newcommand{\IQ}{\mathbb{Q}}
\newcommand{\IH}{\mathbb{H}}
\newcommand{\IR}{\mathbb{R}}
\newcommand{\IC}{\mathbb{C}}
\newcommand{\IF}{\mathbb{F}}
\newcommand{\IV}{\mathbb{V}}
\newcommand{\II}{\mathbb{I}}
\newcommand{\IZ}{\mathbb{Z}}
\newcommand{\re}{{\rm Re}}
\newcommand{\im}{{\rm Im}}
\newcommand{\sym}{{\rm Sym}}

\newcommand{\tmat}[1]{{\tiny \left(\begin{matrix} #1 \end{matrix}\right)}}
\newcommand{\mat}[1]{\left(\begin{matrix} #1 \end{matrix}\right)}
\newcommand{\diff}[2]{\frac{\partial #1}{\partial #2}}
\newcommand{\gen}[1]{\langle #1 \rangle}
\newcommand{\ket}[1]{| #1 \rangle}
\newcommand{\jacobi}[2]{\left(\frac{#1}{#2}\right)}

\newcommand{\drawsquare}[2]{\hbox{%
\rule{#2pt}{#1pt}\hskip-#2pt%  left vertical
\rule{#1pt}{#2pt}\hskip-#1pt%  lower horizontal
\rule[#1pt]{#1pt}{#2pt}}\rule[#1pt]{#2pt}{#2pt}\hskip-#2pt%  upper
horizontal
\rule{#2pt}{#1pt}}% right vertical
\newcommand{\fund}{\raisebox{-.5pt}{\drawsquare{6.5}{0.4}}}
\newcommand{\antifund}{\overline{\fund}}

\newtheorem{theorem}{\bf THEOREM}
\def\thetheorem{\thesection.\arabic{theorem}}
\newtheorem{proposition}{\bf PROPOSITION}
\def\thetheorem{\thesection.\arabic{proposition}}
\newtheorem{observation}{\bf OBSERVATION}
\def\thetheorem{\thesection.\arabic{observation}}

\def\theequation{\thesection.\arabic{equation}}
\newcommand{\setall}{\setcounter{equation}{0}
        \setcounter{theorem}{0}}
\newcommand{\setequation}{\setcounter{equation}{0}}
\renewcommand{\thefootnote}{\fnsymbol{footnote}}

\newcommand{\bq}{\begin{eqnarray*}}
\newcommand{\eq}{\end{eqnarray*}}
\newcommand{\bes}{\begin{subequations}}
\newcommand{\ees}{\end{subequations}}

~\\
\vskip 1cm

\begin{center}
{\Large \bf Gauge Theories and Dessins d'Enfants: \\}
{\Large \bf \qquad Beyond the Torus}
\end{center}
\medskip

\vspace{.4cm}
\centerline{
{\large Sownak Bose}$^{1}$,
{\large James Gundry}$^{2}$ \&
{\large Yang-Hui He}$^{3}$ 
\footnote{sownak.bose@durham.ac.uk; ~jmg202@cam.ac.uk; ~hey@maths.ox.ac.uk}}
\vspace*{3.0ex}

\begin{center}
{\it
{\small
{
${}^{1}$ St. Catherine's College, University of Oxford, OX1 3UJ, Oxford,
UK and \\
Institute for Computational Cosmology, Department of Physics, Durham University, DH1 3LE, Durham, UK\\ 
\vskip 10pt
${}^{2}$ Somerville College, University of Oxford, OX2 6HD, Oxford, UK and \\
Department of Applied Mathematics and Theoretical Physics, University of Cambridge, CB3 0WA, Cambridge, UK \\ 
\vskip 10pt

${}^{3}$ Department of Mathematics, City University, London, EC1V 0HB,
UK and \\
School of Physics, NanKai University, Tianjin, 300071,
P.R.~China and \\
Merton College, University of Oxford, OX1 4JD, UK\\

}
}}
\end{center}

\vspace*{4.0ex}
\centerline{\textbf{Abstract}} \bigskip

{\it Dessin d'Enfants} on elliptic curves are a powerful way of encoding doubly-periodic brane tilings, and thus, of four-dimensional supersymmetric gauge theories whose vacuum moduli space is toric, providing an interesting interplay between physics, geometry, combinatorics and number theory.
We discuss and provide a partial classification of the situation in genera other than one by computing explicit Belyi pairs associated to the gauge theories.
Important also is the role of the Igusa and Shioda invariants that generalise the elliptic $j$-invariant.

\newpage

\tableofcontents

\vspace{2in}

%%%%%%%%%%%%%%%%%%%%%%%%%%%%%%%%%%%%%%%%%%%%%%%%
\section{Introduction}
Over almost a decade, a fruitful programme of investigating certain extraordinary bipartite structure of supersymmetric gauge theories in four-dimensions has emerged. 
What began as a convenient method of encoding the matter content and interactions of world-volume gauge theories of D3-branes probing non-compact Calabi-Yau manifolds that admit a toric description \cite{Feng:2000mi},  has blossomed into a vast field ranging from the field and string theory of configurations of brane tilings \cite{Hanany:2005ve,Franco:2005sm} to the integrable models of dimers \cite{Eager:2011dp,Franco:2012hv}, from the geometry of Calabi-Yau algebras and cluster transformations \cite{Franco:2014nca} to the systematic outlook of bipartite field theories (BCFTs) \cite{Franco:2012mm,Heckman:2012jh,Xie:2012mr} and remarkable relations to scattering amplitudes in $\cN=4$ Super-Yang-Mills theory \cite{ArkaniHamed:2012nw,Golden:2013xva,Cachazo:2012pz,Amariti:2013ija,Franco:2013nwa}.

A particularly enticing direction has been the recasting \cite{Jejjala:2010vb,Hanany:2011ra,Hanany:2011bs,He:2012xw} of the above setup in terms of Grothendieck's {\it dessin d'enfant} \cite{Schneps1994a}, a structure which caused the great master himself  to exclaim: ``I do not believe that a mathematical fact has ever struck me quite so strongly as this one, nor had a comparable psychological impact''. Subsequently, it is natural that a rich interplay between field theory and number theory should emerge \cite{He:2012kw,He:2012js,He:2012jn}.
The concrete realisation of the {\it dessin} is amazingly simple, consisting of a pair -- the so-called Belyi pair -- of data: a Riemann surface $\Sigma$ and a surjective map therefrom unto $\IP^1$.
In explicit coordinates, this is no more than a (hyper)elliptic curve in affine variables $(x,y) \in \IC^2$ and a rational function in $(x,y)$.
What astounded Grothendieck is that this ``very familiar, non-technical nature of the objects considered, of which any child's drawing scrawled on a bit of paper gives a perfectly explicit example'' should encode the subtleties of a holy grail of number theory, the absolute Galois group Gal$(\overline{\IQ}/\IQ)$.
Indeed, the key to the Belyi pair is that the parameters therein are (rigidly) algebraic numbers; while the degree of the field extension over $\IQ$ has been shown to be a Seiberg duality invariant, how these algebraic numbers precisely relate to the R-charges (under isoradial embedding) and to (normalised) volumes in the dual Sasaki-Einstein geometry remains to be understood.

With the aid of modern computing and algorithmic geometry \cite{Hebook}, the combinatorial nature of our theories is especially amenable to a taxonomical analysis, and series of classification results in cataloguing these brane tilings as bipartite graphs and/or as quivers with superpotential has been shedding continual light via experimentation \cite{Hanany:2008fj,Hanany:2008gx,Davey:2009bp,Hewlett:2009bx,Hanany:2011iw,Hanany:2012hi,Hanany:2012vc,Cremonesi:2013aba,He:2012jn,He:2013eqa,He:2014jva}.
Along this vein of thought, it is certainly an important question to write down, and classify where possible, the relevant Belyi pairs.
Unfortunately, this is an extremely difficult task, computationally prohibitive even in seemingly innocuous cases, because we need to solve for the exact roots of high degree polynomial systems. There are only a handful of cases known in the gauge theory literature \cite{Hanany:2011ra}.
Again, with high-power computing and efficient algorithms, there has been encouraging progress \cite{Adrianov2007,Sijsling,Klug,HV,Jones,Elkin}.

While the archetypal brane tilings and dimer models are bipartite graphs on the torus, i.e., {\it dessins} on the elliptic curve, which give us affine Calabi-Yau threefolds, in general, the moduli spaces of gauge theories corresponding to {\it dessins} on genus $g$ Riemann surfaces are Calabi-Yau varieties of dimension $2g+1$ \cite{Cremonesi:2013aba,He:2014jva}.
We will take this comprehensive viewpoint, calculate where needed and make use of the available datasets from the mathematics literature where possible (extensive use will be made of the excellent interactive website of \cite{Elkin}), to explicitly write down the Belyi pairs, genus by genus, and degree by degree.
This catalogue should prove to be valuable to the study of the gauge theories of our concern.
In due course, we will discuss some general strategy in computing Belyi pairs for families of related geometries, as well as the use of invariants beyond the famous Klein $j$-invariant.

The organisation of the paper is as follows. In Section \ref{s:general}, we briefly review and outline the construction of {\it dessins d'enfants} as explicit Belyi pairs of a hyperelliptic curve of arbitrary genus and a rational map therefrom onto $\IP^1$. This structure should encode a supersymmetric gauge theory whose moduli space of vacua is a toric variety.
Next, in Sections \ref{s:g0}, \ref{s:g1} we address genus 0 and 1 respectively.
Genus 1 is the case of the torus, or doubly-periodic brane tiling of the plane and is the most studied class.
We then discuss extensions to higher genera, focusing on the recently studied double-handled tilings of genus 2 in Section \ref{s:g2} and thence, to genus 3 in Section \ref{s:g3}.
We will see how generalisations of the Klein j-invariant, the so-called Igusa and Shioda invariants, are useful in the construction.
We conclude with an outlook in Section \ref{s:conc}.

%%%%%%%%%%%%%%%%%%%%%%%%%%%=================================
\section{Dessins d'Enfants in Arbitrary Genera}\label{s:general}
\setall
In this section, we briefly introduce the main tool in studying the subject of gauge theories in the context of bipartite field theories; namely the theory of {\it dessins d'enfants}. 
The reader is referred to the wonderful books \cite{Schneps1994a,Lando} on {\it dessins} and a rapid introduction and brief review in \cite{He:2012js} for its context in physics.
Simply put, a {\it dessin} is a finite, connected graph, possessing 2-colouring i.e., it is bipartite, with nodes coloured black and white alternately. To this idea, we now apply Belyi's theorem which states that:
%\vspace{5pt}
\begin{thm}If $\Sigma$ is an algebraic curve (complex surface) over $\mathbb{C}$, then it has a model over $\overline{\mathbb{Q}}$ if and only if there exists a holomorphic covering $\beta : \Sigma \longrightarrow \mathbb{P}^1\,\mathbb{C}$, ramified over only three points. These three points may be taken as $\left\{0,1,\infty\right\}$ by a M\"{o}bius transformation.
\end{thm}

We refer to the combination $\left(\Sigma,\beta\right)$ as a {\it Belyi pair}. The Weierstra\ss $\;\wp \left(z\right)$ function allows us to algebraically realise the Riemann surface $\Sigma$ as a hyperelliptic curve of the form 
$
y^2 = f\left(x\right)$,
where the degree of the polynomial $f\left(x\right)$ is such that it is related to the genus of $\Sigma$ as:
\begin{equation}\label{eq:order}
g\;\; \rm{(genus)} \longrightarrow 2g+1\;\; \rm{or} \;\; 2g+2 \;\; \rm{(degree)}.
\end{equation}

The holomorphic map itself is rational, which in its most general form can be written as:
\begin{equation}
\beta\left(x,y\right) = \frac{P\left(x\right)+R\left(x\right)y}{Q\left(x\right)},
\end{equation}
where $P,Q,R$ are polynomials in $x$. Note that any expression in $y^2$ will be reduced to a polynomial in $x$ via the definition of the hyperelliptic curve.

One should know that because two of the preimages can be forced to be at $x=0$ and $x=1$ by means of an $SL(2,\mathbb{C})$ transformation, the curve $y^2=f(x)$ can be written in the factored or so-called Legendre form:
\begin{equation}\label{eq:Legendre}
y^2 = f(x) = x(x-1)(x-\alpha)(x-\beta)(x-\gamma)\dots,
\end{equation}
where once again the order of the polynomial in $x$ is determined by the genus of the surface it describes as in Equation~\ref{eq:order}.
The calculation of the ramification indices follows the methodology set by~\cite{Hanany:2011ra}. It is expedient to introduce the {\it total derivative}, which is the derivative to be used when considering the order of vanishing (i.e., ramification) at the branch points when restricted to $\Sigma$. Defining $F\left(x,y\right) = y^2 - f\left(x\right)$, which must vanish identically on the curve, we have that:

\begin{equation}
\frac{d}{dx} = \frac{\partial}{\partial x} - \frac{\partial_x F}{\partial_y F}\frac{\partial}{\partial y}.
\end{equation}

This expression is valid at the points where $x$ is a good local coordinate i.e., points where the coordinates $\left(x,y\right)$ do not vanish on the curve. This is reflected in the fact that $\partial_y F = 2y$ vanishes at these points and the second term diverges. Therefore, alternatively, we can use:

\begin{equation}
\frac{d}{dy} = \frac{\partial}{\partial y} - \frac{\partial_y F}{\partial_x F}\frac{\partial}{\partial x},
\end{equation}

which is valid when $\partial_x F \neq 0$ and this $y$ is a good local coordinate. Finally, near the point $\left(\infty,\infty\right)$, where a good coordinate is $\epsilon$ with $x = 1/\epsilon^2$ and $y = 1/\epsilon^d$, where $d$ is the degree of the polynomial $f\left(x\right)$, the total derivative can be written as:

\begin{equation}
\frac{d}{d\epsilon} = -2y \frac{\partial}{\partial x} - dx^2\frac{\partial}{\partial y}.
\end{equation}

If $\beta\left(\infty\right) = \infty$, then this derivative is understood to be acting on $1/\beta$. With these in hand, we need only to follow a straightforward routine. If $\left(x_0^i,y_0^i\right)$ is a preimage of 0, then its ramification $r_0\left(i\right)$ is defined where:
\begin{equation}
\left.\frac{d^k}{dx^k}\right|_{\left(x_0^i,y_0^i\right)} \beta\left(x,y\right) = 0,
\end{equation}

%%%%%%%%%%%%%%%%%%%%%%%%%%%%%%%%%%%%
\begin{figure}[t]
\centering
\includegraphics[trim = 0mm 3.5mm 0mm 3.5mm, clip, scale=0.6]{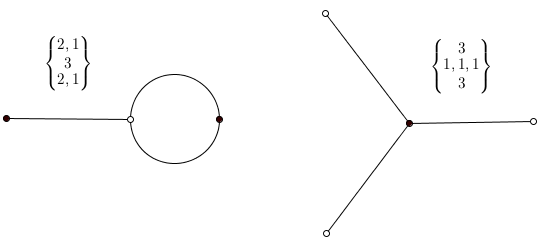}\hfill
\caption{{\sf {\footnotesize Examples of {\it dessins} with the corresponding ramification structures alongside. Note that these examples are purely for illustrative purposes, and not realisable as physical gauge theories, for which we require a node to have at least 2 edges, as well as the ``balanced bipartite conditions" -- i.e., that there are as many white nodes as there are black in the fundamental region of the {\it dessin}.}}}
\label{dessins}
\end{figure}
%%%%%%%%%%%%%%%%%%%%%%%%%%%%%%%%%%%%

for all $k = 0,1,2,\dots,r_0\left(i\right)-1$, where $k=0$ is just evaluation. We then follow a similar procedure to calculate $r_1\left(i\right)$ and $r_\infty(i)$. In order to go from here to the drawing of {\it dessins}, we make the following identifications: for the $m$th preimage of 0, we associate a black node with valency $r_0(m)$ (i.e., $r_0(m)$ edges), and to the $n$th preimage of 1, we associate a white node with valency $r_1(n)$. Given that the {\it dessins} are bipartite, we connect only black nodes to white nodes and vice versa, thereby forming a face, which is a $(2r_\infty(k))-$gon. To compactify the above information for an individual {\it dessin}, represented by the hyperelliptic curve $y^2 = f(x)$, we use the following notation for the {\it ramification structure} (also known as {\bf passport} of the {\it dessin} \cite{Schneps1994a,Lando}):

\begin{equation}
\begin{Bmatrix}
r_0(1),r_0(2),\dots,r_0(B) \\
r_1(1),r_1(2),\dots,r_1(W) \\
r_\infty(1),r_\infty(2),\dots,r_\infty(I) 
\end{Bmatrix}.
\end{equation}
Examples of this notation are shown in Figure~\ref{dessins}. We conclude this section with the Riemann-Hurwitz formula, that allows us to easily translate between the notation above and the genus of the surface $\Sigma$ that it represents:
\begin{equation}\label{eq:RH}
2g - 2 = d - (n_0+n_1+n_\infty),
\end{equation}

where $d$ is the degree of the map $\beta(x,y)$, equal to $\Sigma_i r_0(i) = \Sigma_i r_1(i) = \Sigma_i r_\infty(i)$, and $n_{\left\{0,1,\infty\right\}}$ are the number of ramification points for $\{0,1,\infty\}$ respectively.

In general, a {\it dessin} on a genus $g$ Riemann surface corresponds to a brane tiling whose world-volume physics is an $\mathcal{N}=1$ supersymmetric gauge theory in four dimensions whose (mesonic) moduli space \cite{Cremonesi:2013aba} is an affine toric Calabi-Yau manifold of dimension $2g+1$.
Indeed, for $g=1$, we have the moduli space being a Calabi-Yau threefold: this is why brane tilings on the torus, i.e., doubly-periodic planar tilings are so important in string theory and to AdS/CFT.
In the following sections, we present a discussion on Belyi maps in the case of genus 0, 1, 2 and 3, with reference to specific cases that further illustrate the points made, with focus on the computational aspect of these Belyi pairs.

%%%%%%%%%%%%%%%%%%%%%%%%================================
%%======================================================
\section{Genus 0: Dessins on the Riemann Sphere}
\label{s:g0}
\setall
We will begin by considering the simplest class of toric gauge theories -- those that can be represented as dimer models/brane tilings on a genus 0 Riemann surface. Of course, this space is nothing but the 2-sphere $S^2 \simeq \mathbb{P}^1$. With $g = 0$, a quick check against the Riemann-Hurwitz relation in Equation~(\ref{eq:RH}) tells us that for Belyi pairs in a genus 0 case, the number of ramification points exceeds the degree of the map by 2. An example of the ramification structure would be $\begin{Bmatrix}
3, 2, 2 \\ 3, 2, 2 \\ 3, 2, 2
\end{Bmatrix}$. Clearly in this case, the number of ramification points is 9, while the degree of the map is 7. The fundamental region of this dimer model consists of three black nodes (one with 3 edges, and two with 2), three white nodes (one with 3 edges, and two with 2), and three types of faces ($U\left(N\right)$ gauge groups) -- one with 6 sides, and two with 4 sides.

Because of the fact that the target space is simply a 2-sphere it therefore renders the term ``Belyi pair" a bit of a misnomer in the genus 0 case as there is no need for a hyperelliptic curve. 
Moreover, the Belyi map is simply a rational function $f(x) = P(x) / Q(x)$ in $x$, the projective coordinate of the source $\IP^1$ onto the target $\IP^1$.
Thus, one can imagine a fairly straight-forward algorithm to determine the map and the reader is referred to \cite{Schneps1994a,Lando} and also \cite{Ashok:2006br} for a nice exposition.
For low degree, it is relatively straightforward to calculate $f(x)$ even using software like {\tt MATHEMATICA}, however as the degree goes up and the {\it dessin} becomes more complicated, the algebraic numbers involved quickly grow to formidable complexity, being explicit roots, where possible, of polynomials of high degree.
 
We present a catalogue of ramification structures and their corresponding Belyi maps arising in genus 0 in Appendix \ref{a:g0}. Some maps do not exist, and were ruled out due to the Frobenius relation, thanks to insight from~\cite{Elkin}.
As an explicit illustration, consider the ensuing map which up to the re-definition of $(0,1,\infty)$ gives the trivalent {\it dessin} in the right of Figure \ref{dessins}; note that this shuffling of the critical points makes the {\it dessin} itself look rather different, as we will see below:
\begin{equation}
\begin{Bmatrix}
3 \\ 3 \\ 1,1,1
\end{Bmatrix} \longrightarrow  \frac{\left(-3i + \sqrt{3}\right)x^3}{9\left(i + \sqrt{3} - 2i x\right)\left(x - 1\right)} \;.
\end{equation}
We now record the pre-images of the critical points, the Taylor series around these points whence we can see the ramification index by noticing the lowest power, as well as the corresponding {\it dessin}:
\begin{equation}
\begin{array}{l}
\includegraphics[scale=0.6]{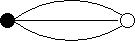}
\end{array}
\qquad
\begin{array}{|c|c|c|c|}\hline
\mbox{image} & \mbox{pre-image} & \mbox{Series}& \mbox{Ram}
\\ \hline \hline
0 & 0 & \frac{\left(-\sqrt{3}+3 i\right) x^3}{9 \sqrt{3}+9 i}+O\left(x^4\right)
& 3 \\
1 & a:= \frac{1}{2} \left(3-i \sqrt{3}\right) & 1+\frac{i (x-a)^3}{3 \sqrt{3}}+O\left((x-a)^4\right) & 3
\\
\infty & 1 & \frac{\frac{1}{6}-\frac{i}{6 \sqrt{3}}}{x-1}+\left(\frac{1}{2}-\frac{i}{6 \sqrt{3}}\right)+O\left((x-1)^1\right) & 1\\
& b := \frac{1}{2} \left(1-i \sqrt{3}\right) &
\frac{3-i \sqrt{3}}{18 (x-b)}+\frac{1}{18} \left(9+i \sqrt{3}\right) +
O\left((x-b)^1\right) & 1\\
& \infty & \frac{1}{18} \left(3+i \sqrt{3}\right)
   x+\frac{1}{3}+O\left(\left(\frac{1}{x}\right)^1\right) & 1\\
\hline
\end{array}
\end{equation}

The above example serves as quite a pedagogical illustration of the type of calculations involved.
Note that all coefficients are algebraic numbers; in the above they are in quadratic extension of $\IQ$.
Of course, in the type of brane-tilings used in the construction of $\cN=1$ world-volume gauge theories of D-branes, we usually do not consider orphan legs - i.e., valency one nodes or mass terms - i.e., valency two nodes - which can be integrated out. In other words, traditionally, only valency 3 and above, corresponding to cubic and higher interaction terms in the superpotential, are considered.
However, due to the study of BFTs in the more general context as discussed in the introduction, we include all valencies and include the relevant {\it dessins} in the catalogue for completeness.
One final point to emphasise in our catalogue is that all the {\it permutations} amongst the three rows of the passport amounts to a linear-fractional transformation among the 3 branch points $(0,1,\infty)$, which can then be composed with the Belyi map presented.

%%%%%%%%%%%%%%%%%%%%%%%%%%==============================
%%%%%%%%%%%%%%%%%%%%%%%%%%==============================
\section{Genus 1: Doubly-Periodic Brane Tilings}
\label{s:g1}
\setall
A reference to the Riemann-Hurwitz formula in Equation~(\ref{eq:RH}) tells us that in the genus 1 case, the number of ramification points is equal to the degree of the map. The canonical example used to illustrate this case is that of $\mathbb{C}^3$, the {\it dessin} for which is shown below. It should be noted that the map locally looks like $w = z^3$ around the 3 ramification points.
Recall that the permutation triple has $\sigma_{B}$, $\sigma_{W}$,
and $\sigma_{\infty}$ all equal to $(123)$. This means that the
ramification structure is $\begin{Bmatrix}3\\
3\\
3
\end{Bmatrix}$. This notation means that we require zero, one, and infinity to each
have only one preimage on the torus, and that the ramification indices
of these points must be three. So we require the map to look like
$w=z^{3}$ in local coordinates at these three points. The Belyi Pair
with these properties can be written
\begin{equation}\label{eq:BC}
\beta(x,\, y)=\frac{1}{2}(1+y)\qquad,\qquad y^{2}=x^{3}+1\,\,.
\end{equation}

\noindent
We can briefly exhibit the ramification structure of this pair. The
preimage of zero is found by solving $\beta(x,\, y)=0$, which gives
$y=-1$ and so $x=0$ (from consulting the torus). We wish to see
what the good local coordinates are around the point $(0,-1)$, and
so we substitute $\begin{array}{c}
x=0+\delta x\end{array}$ and $y=-1+\delta y$ into the torus. Taking only the leading order
in the small quantities $\delta x$ and $\delta y$, we find that
$\delta y\sim\delta x^{3}$ and so we can take $\delta y\sim\epsilon^{3}$
as a good local coordinate. Substituting $y=-1+\epsilon^{3}$ into
$\beta(x,\, y)$ shows that the ramification index of our (only) preimage
of zero is indeed three. The preimages of one and infinity are similarly
structured.
\begin{figure}[t]
\centering
\includegraphics[trim = 0mm 3.5mm 0mm 3.5mm, clip, scale=0.6]{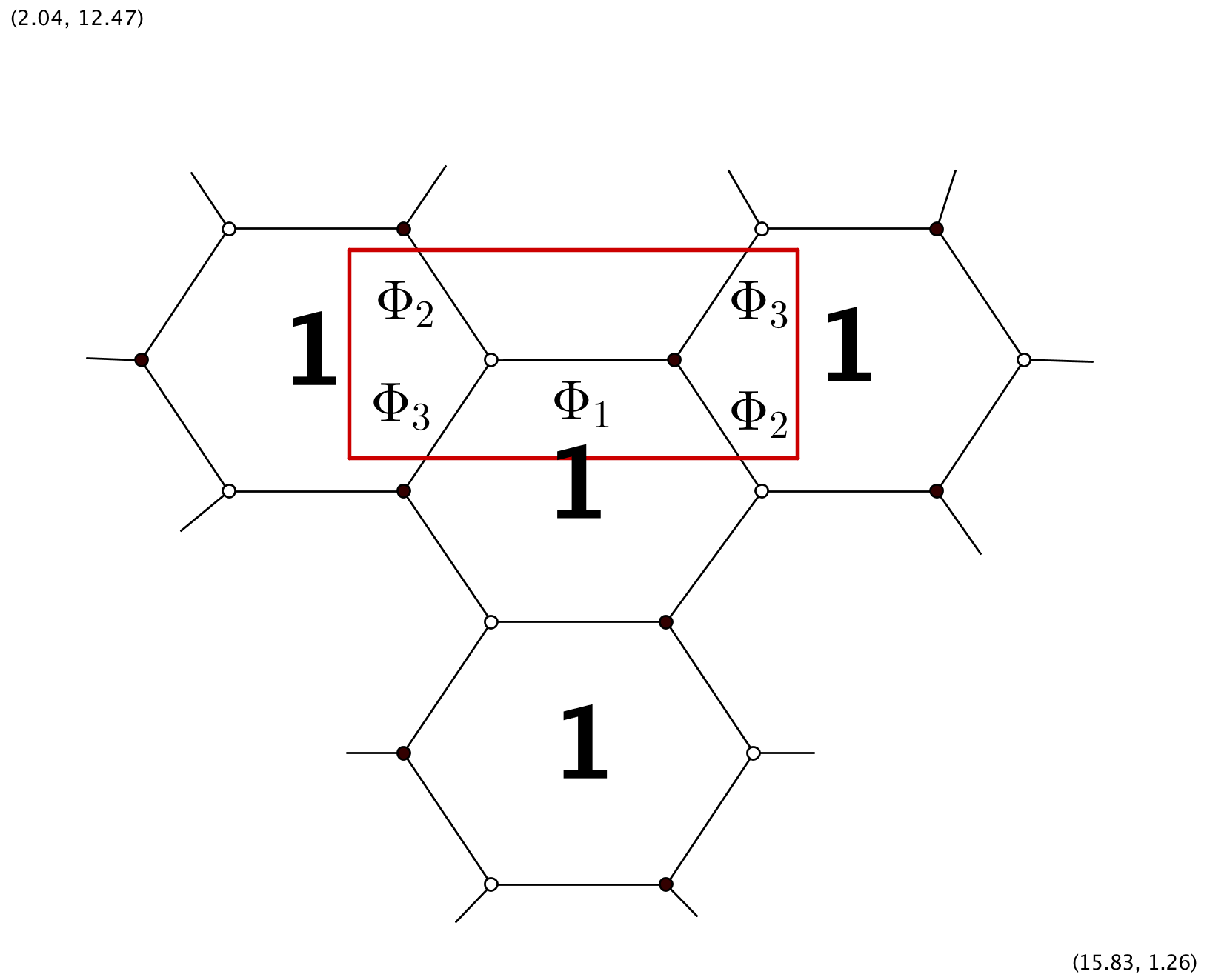}
\caption{\sf {\footnotesize Dimer for the $\mathbb{C}^3$ theory. The red section encloses the so-called ``fundamental region" of the dimer. This fundamental region is then periodically tiled over the surface of the torus on which it is embedded, and it is this object that we call the {\it dessin d'enfant}. We label the edges in terms of fields to be included in the superpotential for our theory, whereas the hexagonal face represents the relevant U$\left(N\right)$ gauge group. The label ``1" refers to the fact that there is only one gauge group in the case of $\mathbb{C}^3$.}}
\label{C3dimer}
\end{figure}

%%%%%%%%%%%%%%%%%%%%%%%
\subsection{$j-$invariants and Coordinate Transformations}
\label{jinvsec}
Of course, the Belyi pair presented in Equation~(\ref{eq:BC}) is by no means unique. We can always find a coordinate transformation in $\left(x, y\right)$ that will map our current Belyi pair to a new one. The concept of $j-$invariants allows us to identify sets of elliptic curves that are equivalent in the sense that they represent the same toric surface, and vary only by some coordinate transformation. For an elliptic curve defined as 
\begin{equation}\label{eq:ecurve}
y^2 = 4x^3 - g_2 x - g_3
\ ,
\end{equation}
the $j-$invariant~\cite{Megyesi} is defined as:
\begin{equation} \label{eq:jinv}
j = 1728 \frac{g_2^3}{g_2^3 - 27g_3^2}\;.
\end{equation}
For example, one alternative representation of $\mathbb{C}^3$~\cite{Hanany:2011ra} is the Belyi pair:
\begin{equation}
y^2 = x\left(x - 1\right)\left(x - \frac{1}{2}\left(1 - i \sqrt{3}\right)\right) , \;\;\;\;\;\; \beta\left(y\right) = \frac{1}{2}\left(1 + \left(-1\right)^{\frac{1}{4}}3^{\frac{3}{4}}y\right)\;.
\end{equation}
Despite the fact that the ramification structure is exactly as it should be for $\mathbb{C}^3$, the Belyi pair itself looks completely different to that in Equation~(\ref{eq:BC}). A verification of the $j-$invariant for each of the elliptic curves yields $j = 0$ for both of the cases, indicating that the two curves represent the same toric surface, related by some coordinate transformation. While neither $y^2 = x^3 + 1$ nor $y^2 = x\left(x - 1\right)\left(x - \frac{1}{2}\left(1 - i \sqrt{3}\right)\right)$ are in the form given in Equation~(\ref{eq:ecurve}), a theorem due to Nagell allows us to transform any cubic curve to the form $y^2 = 4x^3 - g_2 x - g_3$, validating our use of Equation~(\ref{eq:jinv}) for the $j-$invariant (see~\cite{Nagell1929} for more details). The transformation in this case $\left\{x \rightarrow \frac{2\left(x + 1\right)}{3 + i\sqrt{3}}, y \rightarrow \left(-1\right)^{\frac{-1}{4}}3^{\frac{-3}{4}}y\right\}$ takes the above Belyi pair to the standard form shown in Equation~(\ref{eq:BC}).

%%%%%%%%%%%%%%%%%%%%%%%%%%%%%%%
\subsection{Fertile Elliptic Curves}
In this section we present several appealing parameterisations of elliptic
curves. The table following each curve provides details of its noteworthy
points. Whilst all elliptic curves can be brought by coordinate transformation
into the Weiestra\ss\,form (Equation~\ref{eq:Legendre}), when thinking through the construction
of Belyi Pairs one of these different parameterisations
may be more instructive because the points with nontrivial orders
of vanishing in the good local coordinate $\epsilon$ are exhibited
explicitly. Generic points (i.e. those with no special properties)
are represented as $(G,\, g)$. The columns headed by $\delta x$
and $\delta y$ contain the variation of $x$ and $y$ in terms of
the good local coordinate $\epsilon$, whilst the multiplicity columns
state how many points on the torus are specified by choosing that
particular value of $x$ or $y$. The cube roots of unity are written
as $\omega_{i}=(1,\, e^{i\frac{2\pi}{3}},\, e^{i\frac{4\pi}{3}})$.

\newpage
%%%%%%%%%%%%%%%%%%%%%%%%%%%%%%%%%%%%%%%%%%%%%%%%%%%%%%%%%
\subsubsection{$y^{2}=x(x-1)(x-\lambda)$}

\noindent \begin{center}
\vspace{-1.0em}%
\begin{tabular}{|c|c|c|c|c|c|}
\hline 
$x$ & $y$ & $\delta x$ & $\delta y$ & $x$ - multiplicity & $y$ - multiplicity\tabularnewline
\hline 
\hline 
$G$ & $g$ & $\epsilon$ & $\epsilon$ & $2$ & $3$\tabularnewline
\hline 
$0,\,1,\,\lambda$ & $0$ & $\epsilon^{2}$ & $\epsilon$ & $1$ & $3$\tabularnewline
\hline 
$\infty$ & $\infty$ & $\epsilon^{-2}$ & $\epsilon^{-3}$ & $1$ & $1$\tabularnewline
\hline 
\end{tabular}
\par\end{center}

This form is the Legendre Normal Form set out in \cite{Hanany:2011ra}.

%%%%%%%%%%%%%%%%%%%%%%%%%%%%%%%%%%%%%%%%%%%%%%%%%%%%%%%%%%%
\subsubsection{$(y-a)^{2}=(x-\gamma)^{3}+\eta^{2}$} 

\noindent \begin{center}
\vspace{-0.8em}%
\begin{tabular}{|c|c|c|c|c|c|}
\hline 
$x$ & $y$ & $\delta x$ & $\delta y$ & $x$ - multiplicity & $y$ - multiplicity\tabularnewline
\hline 
\hline 
$G$ & $g$ & $\epsilon$ & $\epsilon$ & $2$ & $3$\tabularnewline
\hline 
$\gamma$ & $a\pm\eta$ & $\epsilon$ & $\epsilon^{3}$ & $2$ & $1$\tabularnewline
\hline 
$\gamma-\omega_{i}\eta^{ \frac{2}{3}}$ & $a$ & $\epsilon^{2}$ & $\epsilon$ & $1$ & $3$\tabularnewline
\hline 
$\infty$ & $\infty$ & $\epsilon^{-2}$ & $\epsilon^{-3}$ & $1$ & $1$\tabularnewline
\hline 
\end{tabular}
\par\end{center}

This parameterisation is particularly appealing
due to the appearance of the points which vanish with a cubic dependence
on the good local coordinate, $\delta y\sim\epsilon^{3}$.

%%%%%%%%%%%%%%%%%%%%%%%%%%%%%%%%%%%%%%%%%%%%%%%%%%%%%%%%%%%
\subsubsection{$(y-a)(y-b)(y-c)=(x-\gamma)^{3}$}

\noindent \hspace{-2.5em}%
\begin{tabular}{|c|c|c|c|c|c|}
\hline 
$x$ & $y$ & $\delta x$ & $\delta y$ & $x$ - multiplicity & $y$ - multiplicity\tabularnewline
\hline 
\hline 
$G$ & $g$ & $\epsilon$ & $\epsilon$ & $3$ & $3$\tabularnewline
\hline 
$\gamma$ & $a,\, b,\, c$ & $\epsilon$ & $\epsilon^{3}$ & $3$ & $1$\tabularnewline
\hline 
$\eta${*} & $\frac{1}{3}(a+b+c\pm\sqrt{a^{2}+b^{2}+c^{2}-(ab+bc+ca)})$ & $\epsilon^{2}$ & $\epsilon$ & $2$ & $3$\tabularnewline
\hline 
$\infty$ & $\infty$ & $\epsilon^{-1}$ & $\epsilon^{-1}$ & $1$ & $1$\tabularnewline
\hline 
\end{tabular}

{*}Here, $\eta$ represents six values of $x$. Each of these provides
one of the values of $y$ shown, with $\delta x\sim\epsilon^{2}$,
along an extra trivial value of $y$ which has $\delta x\sim\epsilon$.

The advantage of this elliptic curve is that we have an additional
value of $y$ with $\delta y\sim\epsilon^{3}$, but we have obtained
it at the cost of losing the useful $\delta y\sim\epsilon^{-3};\,\delta x\sim\epsilon^{-2}$
vanishing orders at infinity.

%%%%%%%%%%%%%%%%%%%%%%%%%%%%%%%%%%%%%%%%%%%%%%%%%%%%%%%%%%%
\subsubsection{$(y-a)(y-b)=(x-\gamma)(x-\eta)^{2}$}

\noindent \begin{center}
\vspace{-0.8em}%
\begin{tabular}{|c|c|c|c|c|c|}
\hline 
$x$ & $y$ & $\delta x$ & $\delta y$ & $x$ - multiplicity & $y$ - multiplicity\tabularnewline
\hline 
\hline 
$G$ & $g$ & $\epsilon$ & $\epsilon$ & $2$ & $3$\tabularnewline
\hline 
$\gamma$ & $a,\, b$ & $\epsilon$ & $\epsilon$ & $2$ & $2$\tabularnewline
\hline 
$\eta$ & $a,\, b$ & $\epsilon$ & $\epsilon^{2}$ & $2$ & $2$\tabularnewline
\hline 
$\infty$ & $\infty$ & $\epsilon^{-2}$ & $\epsilon^{-3}$ & $1$ & $1$\tabularnewline
\hline 
\end{tabular}
\par\end{center}
\noindent
The reader should note that in each of these elliptic curves (except
the first) we have tried to maximise the number of free parameters.
This is because, when trying to construct Belyi Pairs it is necessary
to adjust parameters in order to obtain the desired ramification structure
(which dictates the gauge theory). Hence, having as many as possible
to play with helps to make the task easier. Whilst in principle many
of these parameters are removable by coordinate transformation, the
transformation would alter the map $\beta(x,\, y)$, and may ``damage''
the ramification structure exhibited by the pair. This ramification
structure encodes the gauge theory, and so allowing our elliptic curve
to have free parameters enables us to fine tune the torus and map
to fit the gauge theory.

%%%%%%%%%%%%%%%%%%%%%%%%
\subsection{Combinations and Reformations of Belyi Maps}

Given the computational complexity in constructing explicit Belyi airs,
it is expedient to see whether we can obtain new pairs given simpler ones.
In \cite{Jejjala:2010vb,Hanany:2011ra}, the situation of geometric orbifolds 
by Abelian groups was addressed. There, one simply applies an unbranched cover 
of the torus. In the ensuing, we will consider how some algebraic manipulations on 
the underlying curve and on the Belyi map generates new theories.

\subsubsection{Combinations}

In this subsection we will set out the rules for combining two maps
$\beta_{1}$ and $\beta_{2}$ defined on the same elliptic curve.
We will describe the ramification structure of the new map $\beta(\beta_{1},\,\beta_{2})$
in terms of the ramification structures of $\beta_{1}$ and $\beta_{2}$.
The combination we will consider is the product of two maps, $\beta=\beta_{1}\beta_{2}$.

We will deal first with the preimages of zero and infinity before
discussing the more troublesome preimages of one. The reader is reminded
that it is the ramification indices of these points which encode the
gauge theory.

Provided $\beta_{1}^{-1}(0)$ and $\beta_{2}^{-1}(0)$ do not coincide
then the new ramification structure at zero possesses the points of
both constituent ramification structures, with the same indices as
they previously held. If they do coincide then we must add the corresponding
ramification indices. The ramification structure of infinity works
the same way. If a preimage of zero coincides with a preimages of infinity then
the two may cancel out to some order. We find the resulting ramification
index of zero by subtracting the infinity index from the zero index. This is more easily seen if we abandon generality and try out an
example. For our example we will take the $\mathbb{C}^{3}$ Belyi
Pair, and a somewhat arbitrary map for $\beta_{2}$:
\begin{equation}
\label{C3diff}
\beta_{1}=\frac{1}{2}(1+y)\qquad,\qquad\beta_{2}=\frac{e^{i \frac{\pi}{6}}}{\sqrt{3}}(1+x)\,\,.
\end{equation}

The maps $\beta_{1}$ and $\beta_{2}$ will both be defined upon the
same curve $y^{2}=x^{3}+1$. The ramification structures of these
maps are $\begin{Bmatrix}3\\
3\\
3
\end{Bmatrix}$ and $\begin{Bmatrix}2\\
2\\
2
\end{Bmatrix}$ respectively. It should be noted that $\beta_{2}$ on the curve $y^{2}=x^{3}+1$
is not a satisfactory Belyi Pair, because the number of ramification
points, three, is not equal to the degree of the map, two. By the
Riemann-Hurwitz formula, this does not correspond
to a map from a curve of genus one, and hence is not a map from a
torus. Nevertheless it is worth considering non-Belyi Pairs if they
are intermediate steps in constructing Belyi Pairs. We now wish to establish the ramification structure of the new map
$\beta=\beta_{1}\beta_{2}$. Let us consider first the preimages of
zero:
\[
\beta_{1}^{-1}(0)=(0,-1)\qquad,\qquad\beta_{2}^{-1}(0)=(-1,\,0)\,\,.
\]

These clearly do not coincide, so we conclude (by adding together
the previous ramification structures) that the new ramification structure
of zero, for the map $\beta$, is%
\footnote{This notation simply isolates one row of the ramification structure,
with the first digit dictating the isolated row.%
} $0:\{23\}$. Next we should consider the preimages of infinity. For
both curves these are at $(\infty,\,\infty)$. Here we have a case
in which the preimages clearly do coincide, and so we combine the
``two'' ramification points into one, and add together their ramification
indices. This tells us that the new ramification structure of infinity
is $\infty:\{5\}$.

The ramification structure of one is considerably more awkward to
deal with. In general, the previous two ramification structures are
destroyed, and we are left with a string of ``trivial'' ramification
points, giving ramification structure $1:\{111...\}$. Forcing the
preimages of one to provide an interesting ramification structure
is what makes the construction of Belyi Pairs difficult.

To locate the preimages of one, we must solve simultaneously
\[
\beta_{1}(x,\, y)\beta_{2}(x,\, y)=1\qquad,\qquad y^{2}=x^{3}+1\,\,.
\]

For the example under consideration we obtain five solutions, each
of trivial ramification. This tells us that the new ramification structure
of one is $1:\{11111\}$, and then we can say that the overall ramification
structure of $\beta=\beta_{1}\beta_{2}$ is $\begin{Bmatrix}23\\
11111\\
5
\end{Bmatrix}$. This is not a Belyi Pair, since the degree of the map, five, is
not equal to the number of ramification points, eight, which is required
by the Riemann-Hurwitz relation for a map from a curve of genus one
\cite{Jejjala:2010vb}.

The two maps we started with in this example contained no free parameters,
and so there was no room for adjusting the new preimages of one. Without
careful adjusting of parameters, the only way in which the ramification
structures of one can combine nontrivially is if the preimage of a
complex number $\alpha$ via $\beta_{1}$ coincides with the preimage
of $\alpha^{-1}$ via $\beta_{2}$. Then we obtain a combined ramification
point, with ramification index equal to the smaller index out of the
two.

An example of this (for $\alpha=1$) can be seen when combining $\beta_{1}=\frac{1}{2}(1+y)$
and $\beta_{2}=(1+x^{2})$ on the same curve as previously. These
maps both have a preimages of one at the point $(0,\,1)$. These combine
into a single ramification point, of ramification index $2$, which
is the smaller index, originating from $\beta_{2}$.

%%%%%%%%%%%%%%%%%%%%%%%%%%%%%%%%%%%%%%%%%%%%%%%%%%%%%%%%%%%%%%%%%%%%%%%
\subsubsection{Reformations}

In this subsection we present several possible reformations of Belyi
Maps, i.e. simple functions of a Belyi Map $\tau(x,\, y)$ which shuffle
the preimages to give a new map $\beta(x,\, y)$. This may be useful
if, for example, we wish to adjust parameters to fix the preimages
of infinity rather than of one. In that case we would use the reformation
$\frac{1}{1-\tau}$. The tables give $\beta$ as a function of the
old map $\tau$, and indicate how the preimages have been reshuffled.

\noindent \vspace{0.3em}

\noindent \hspace{-1.8em}%
\begin{tabular}{cc}
\begin{tabular}{|c|c|c|c|}
\hline 
$\beta(\tau)$ & $\tau(\beta=0)$ & $\tau(\beta=1)$ & $\tau(\beta=\infty)$\tabularnewline
\hline 
\hline 
\multirow{1}{*}{$\frac{1}{\tau}$} & $\infty$ & $1$ & $0$\tabularnewline
\hline 
$1-\tau$ & $1$ & $0$ & $\infty$\tabularnewline
\hline 
$\frac{1}{1-\tau}$ & $\infty$ & $0$ & $1$\tabularnewline
\hline 
$\frac{1}{2}(\tau+\frac{1}{\tau})$ & $\pm i$ & $1$ & $0\,,\,\infty$\tabularnewline
\hline 
\end{tabular} & %
\begin{tabular}{|c|c|c|c|}
\hline 
$\beta(\tau)$ & $\tau(\beta=0)$ & $\tau(\beta=1)$ & $\tau(\beta=\infty)$\tabularnewline
\hline 
\hline 
\multirow{1}{*}{$\frac{\tau+1}{\tau-1}$} & $-1$ & $\infty$ & $1$\tabularnewline
\hline 
$\frac{1+\tau}{1-\tau}$ & $-1$ & $0$ & $1$\tabularnewline
\hline 
$\frac{1}{2}(1+\tau)$ & $-1$ & $1$ & $\infty$\tabularnewline
\hline 
$\frac{\tau-1}{\tau}$ & $1$ & $\infty$ & $0$\tabularnewline
\hline 
\end{tabular}\tabularnewline
\end{tabular}

%%%%%%%%%%%%%%%%%%%%%%%%%%%%%%%%%%%%%%%%%%%%%%%%%%%%
\subsection{Philobelyiical Investigations}
In this section we present some short investigations regarding the
ramification structures which can be obtained from specific ansätze.
The elliptic curves referred to are those exhibited in Section 4.2,
and we will make use of the results from Section 4.3.

%%%%%%%%%%%%%%%%%%%%%%%%%%%%%%%%%%%%%%%%%%%%%%%%%%%%%
\subsubsection{$\beta(x,\, y)=\alpha x$}

We will consider the map $\beta=\alpha x$ acting on the elliptic
curve from section 4.2.2, $(y-a)^{2}=(x-\gamma)^{3}+\eta^{2}$, and
deduce its possible ramification structures. The parameter $\alpha$
is a non-zero complex number. Starting with zero, we see that $x=0$
is required. This is a generic point with $x$-multiplicity two, and
so without tuning the parameters the ramification structure of zero
is $0:\{11\}$. However, we can alter this by tuning the parameters.
If we set $\eta^{2}=\gamma^{3}$ then $x=0$ is no longer generic
and instead we have a nontrivial $y=a$ point, with $\delta x\sim\epsilon^{2}$.
With this tuning the new ramification structure of zero is $0:\{2\}$.
In a Belyi Pair the reader is reminded that this structure would correspond
to a black node in the dimer with two connecting edges, i.e. two superfields in the gauge theory.

The ramification structure of infinity is easy to deduce - the only
preimage of infinity is the point $(\infty,\,\infty)$ on the elliptic
curve. Around this point we have $\delta x\sim\epsilon^{-2}$, and
so the ramification structure of infinity is $\infty:\{2\}$. In this
case there can be no tuning of parameters to alter this structure.

The preimages of one are easily seen to be at $x=\alpha^{-1}$. Again,
without tuning, this is a generic point and the ramification structure
is $1:\{11\}$, but as with zero we can make the ramification structure
more interesting by setting $\alpha^{-1}=\gamma(1-e^{i \frac{2\pi}{3}})$.
This tuning forces the preimage of one to be a $y=a$ point and so
we obtain $\delta x\sim\epsilon^{2}$ and hence a white node in the
dimer with two connecting superfields, $1:\{2\}$.

With the tuning described then, the ansatz $\beta=\alpha x$ on the
curve (4.2.2) has the ramification structure $\begin{Bmatrix}2\\
2\\
2
\end{Bmatrix}$. This is not a Belyi Pair, but this example has shown how a nontrivial
ramification structure can be obtained from a very simple ansatz by
the action of tuning the parameters of a fertile elliptic curve.

%%%%%%%%%%%%%%%%%%%%%%%%%%%%%%%%%%%%%%%%%%%%%%%%%%%%%%%%%%%%%%%%%%%%%
\subsubsection{$\beta(x,\, y)=\alpha(\frac{y-\mu}{y-\nu})$}

This simple ansatz is steeped in possibilities. Let us consider it
upon the curve (4.2.2). Without tuning, the preimages of zero, $y=\mu$,
are three generic points, giving us a ramification structure of $0:\{111\}$.
The preimages of infinity and one are similarly generic, at $y=\nu$
and $y=\frac{\alpha\mu-\nu}{\alpha-1}$ respectively. Without tuning
then, the ramification structure is $\begin{Bmatrix}111\\
111\\
111
\end{Bmatrix}$. There are several paths we can now consider. The first case is to
choose $\begin{pmatrix}\alpha\\
\mu\\
\nu
\end{pmatrix}=\begin{pmatrix}1\\
a-\eta\\
a+\eta
\end{pmatrix}$. By consulting the table for curve (4.2.2) we see that this makes the
preimages of zero, one, and infinity all have $\delta y\sim\epsilon^{3}$.
The ramification structure is then $\begin{Bmatrix}3\\
3\\
3
\end{Bmatrix}$. We have then, with remarkable ease, derived a Belyi Pair for the
$\mathbb{C}^{3}$ theory. Note that the parameters $a,\,\eta,$ and
$\gamma$ are still free. Provided the choice does not disrupt the
ramification structure (e.g. $\eta=0$ would set $\beta(x,\, y)=1$)
then any choice of these parameters is a Belyi Pair for $\mathbb{C}^{3}$.
It is important to note though, that this is not a different Belyi
Pair to the one given in section 2 (from \cite{Hanany:2011ra}) because
the two can be related to each other by coordinate transformation. It follows that there are many ways of writing the Belyi Pair for
a given theory. Simplicity must be traded off between the map and
the elliptic curve. For example, we can express $\mathbb{C}^{3}$
with a simple map $\beta=y$, provided the elliptic curve is the somewhat
less elegant $y(y-1)=x^{3}$. As in the case discussed earlier
in Eq.~\ref{C3diff}, this is another coordinate-transformed way of writing the Belyi
pair for $\IC^3$.

A second path will lead us to the creation of a further Belyi Pair.
If instead of the previous tuning for $\mu$ and $\nu$ we leave these
as generic (though retain $\alpha=1$) then we will, before further
action, have the ramification structure $\begin{Bmatrix}111\\
3\\
111
\end{Bmatrix}$. In accordance with the procedure laid out in Section 4.3 we can combine
this map with two more copies of itself, and hence consider its cube.
The map is now
\[
\tilde{\beta}=\left(\frac{y-\mu}{y-\nu}\right)^{3}\,\,.
\]

The results of section 4.3.1 tell us that the new ramification
structure is $\begin{Bmatrix}333\\
3\,?\\
333
\end{Bmatrix}$, where we have indicated with a question mark that the remainder of the ramification structure
of one is to be deduced. We find that the finite preimages of one
are comprised of two value of $y$:
\[
y(\tilde{\beta}=1)=\frac{\mu-\nu\omega}{1-\omega},\,\frac{\mu-\nu\omega^{2}}{1-\omega^{2}}\qquad;\qquad\omega=e^{i \frac{2\pi}{3}}\,\,.
\]

Without further tuning this would give us $\begin{Bmatrix}333\\
3\,111\,111\\
333
\end{Bmatrix}$, but if these two points can be tuned to $y=a-\eta$ and $y=a+\eta$
then we will obtain the nontrivial order of vanishing $\delta y\sim\epsilon^{3}$.
This is found to occur when we set $\begin{pmatrix}\mu\\
\nu
\end{pmatrix}=\begin{pmatrix}a-\eta(1+2\omega)\\
a-\eta(1+2\omega^{2})
\end{pmatrix}$. With this tuning the ramification structure becomes $\begin{Bmatrix}333\\
333\\
333
\end{Bmatrix}$. The map and curve now constitute a Belyi Pair for the so-called
$dP_{0}$ theory. Unfortunately this pair is found to be related by
coordinate transformation to the one given in \cite{Jejjala:2010vb} for
this theory, and so is not new. Nevertheless it is encouraging that
a Belyi Pair with a complicated ramification structure can be generated
in such a straightforward manner by considering simple ansätze and
the rules for combining them.

%%%%%%%%%%%%%%%%%%%%%%%%%%%%%%%%%%%%%%%%%%%%%%%%%%%%%%%%%%
\subsection{A New Belyi Pair - $PdP_{4}$}

The ansatz from the previous subsection has even more to give us.
Let us consider it with $\begin{pmatrix}\alpha\\
\mu\\
\nu
\end{pmatrix}=\begin{pmatrix}1\\
a-\eta\\
a+\eta
\end{pmatrix}$ again, so that the ramification structure is $\begin{Bmatrix}3\\
3\\
3
\end{Bmatrix}$. Using the procedure laid out in Section 4.3.1 we now wish to combine
this map with a new copy of the ansatz (with new arbitrary parameters
$\mu'$ and $\nu'$), which has been raised to the fourth power:
\[
\beta(x,\, y)=\left(\frac{y-(a-\eta)}{y-(a+\eta)}\right)\left(\frac{y-\mu'}{y-\nu'}\right)^{4}\qquad,\qquad(y-a)^{2}=(x-\gamma)^{3}+\eta^{2}\,\,.
\]

Without making any adjustment of parameters the ramification
structure is now $\begin{Bmatrix}3\,444\\
3\,111\,111\,111\,111\\
3\,444
\end{Bmatrix}$, as described in Section 4.3.1. If this is to become a Belyi Pair then
the ramification structure of one is clearly in need of some work.
The preimages of one are found to be solutions to the equation
\begin{equation}
(y-(a-\eta))(y-\mu')^{4}-(y-(a+\eta))(y-\nu')^{4}=0\,\,.
\end{equation}

In general this has four solutions in $y$, and since the $y$-multiplicity
of generic points on this curve is three, we get twelve trivial points.
However this would not be the case if the left hand side of equation
(4.13) were to factorise nontrivially. In
particular, we would find it useful if, for some choice of the free
parameters, we could force for all $y$ the equality
\begin{equation}
(y-(a-\eta))(y-\mu')^{4}-(y-(a+\eta))(y-\nu')^{4}=\sigma(y-\phi)^{2}(y-\xi)^{2}\,\,.
\end{equation}

After a combination of pen-and-paper algebra and {\tt MATHEMATICA}
computation we find that this equality is indeed possible. The expressions
are cumbersome and so we will not reproduce them here, but it is worth
noting that if we are careful we are left with the parameters $\mu'$,
$\nu'$, and $\gamma$ free. A convenient choice for these parameters
is $\begin{pmatrix}\mu'\\
\nu'\\
\gamma
\end{pmatrix}=\begin{pmatrix}1\\
-1\\
0
\end{pmatrix}$. The $j$-invariant of the elliptic curve is zero (see appendix C)
and the pair are now expressed as
\[
\beta(x,\, y)=\left(\frac{y(1+i)+1}{y(1+i)-1}\right)\left(\frac{y-1}{y+1}\right)^{4}\qquad,\qquad y^{2}=x^{3}-\frac{i}{2}\,\,.
\]

The preimages of one are now $(\infty,\,\infty)$ and the six generic
points described by $y=\phi$ and $y=\xi$. The ramification structure
is $\begin{Bmatrix}3\,444\\
3\,222\,222\\
3\,444
\end{Bmatrix}$. We are tantalisingly close to a new Belyi
Pair. The number of ramification points is equal to the degree of
the map, fifteen. The only remaining problem is that all the gauge
theories which we wish to describe have dimers with equal numbers
of black and white nodes. Clearly our ramification structure has three
more white nodes than it has black nodes. Fortunately we can use a
result from section 4.3.2 and act with the reformation $\beta=\frac{1}{1-\tau}$
on our map to shuffle the preimages. Our final Belyi Pair is now expressed
as \vspace{0.5em}
\begin{equation}
\beta(x,\, y)=\frac{(y+1)^{4}(y(1+i)-1)}{2(y^{2}(2+i)+i)^{2}}\qquad,\qquad y^{2}=x^{3}-\frac{i}{2}\,\,.
\end{equation}

The ramification structure is now $\begin{Bmatrix}3\,444\\
3\,444\\
3\,222\,222
\end{Bmatrix}$, and (4.15) constitutes an original Belyi Pair. This corresponds to the
so-called $PdP_{4}$ theory, or pseudo del Pezzo 4, a toric Calabi-Yau cone over a special del Pezzo surface of degree 5, first introduced in \cite{Feng:2002fv}.

%\vspace{-0.5em}
%%%%%%%%%%%%%%%%%%%%%%%%================================
%%======================================================
\section{Genus 2: Doubly-Handled Tilings}
\label{s:g2}
\setall
We are now in a position to study Belyi pairs (and by extension, gauge theories) that can arise in the case of a dimer model embedded on a genus 2 Riemann surface, which have recently been studied in \cite{Cremonesi:2013aba, He:2014jva}. In this section, we will construct and present a new Belyi pair, the simplest one arising in a genus 2 case -- that of $\begin{Bmatrix}
5 \\ 5 \\ 5
\end{Bmatrix}$. This is similar to the $\mathbb{C}^3$ case analysed in the genus 1 scenario, in that the fundamental region for this dimer contains only one white node and one black node. The difference now is that each node has 5 associated edges (fields), and the solitary face (gauge group) is now a $2 \times 5 = 10-$sided polygon.

%%%%%%%%%%%%%%%%%%%%%%%%%%%%%%%%%%%%%%%%%%%%
\subsection{Explicit Construction}
Our starting point is to first verify that this is indeed a valid ramification structure in a a genus 2 case, and for this, we turn as before to the Riemann-Hurwitz formula (Equation~(\ref{eq:RH})). Using now that $g = 2$, and the number of points as 3, we see that the degree of the map should be 5, as we have above. Once more with $\mathbb{C}^3$ as our motivation, let us use as an ansatz the following pair:
\begin{equation}
\left(y - \alpha\right)^2 = x^5 +1, \;\;\;\;\;\; \beta = \beta\left(y\right)\;.
\end{equation} 
In the above, we have maintained a simple, but more general form for the (hyper)elliptic curve (with $\alpha$ as some complex parameter), adapted to genus 2 by changing the order of the polynomial in $x$ to 5, the order of the map. We have also started off with assuming that the Belyi map $\beta$ is a function of $y$ only. We will see that this, while a simplistic assumption, is enough to carry us through to the end.\\\\
The ramification structure requires that there is only one black node -- that is, only one preimage for 0 when mapped onto the Riemann Sphere. As such, we know that solutions to the equation $\beta\left(y\right) = 0$ must vanish on the curve ($x = 0$). This gives us
\begin{equation}
\begin{split}
\left(y - \alpha\right)^2 = 0 + 1 
\Rightarrow y = \pm 1 + \alpha\;.
\end{split}
\end{equation}
We are free to choose either solution, so taking the negative one reveals a more explicit form for the map $\beta\left(y\right) = A\left(y + 1 - \alpha\right)$, where $A$ is some overall complex factor. At this stage, a quick verification of the ramification structure yields $\begin{Bmatrix}
5 \\ 1,1,1,1,1 \\ 5
\end{Bmatrix}$, which is not a Belyi pair, since it violates our preset balanced bipartite graph condition (that we want the same number of black and white nodes -- hence the same number of preimages for 0 and 1). In order to tweak the preimages of 1 down to only a single ramification point, we can make use of the undermined factor $A$. We want to solve the case of $\beta\left(y\right) = A\left(y + 1 - \alpha\right) = 1$ for the case where the preimage once again vanishes on the curve, so we have
\begin{equation}
\begin{split}
\beta\left(y\right) = A\left(y + 1 - \alpha\right) = 1 
\Rightarrow y = \frac{1}{A} - 1 + \alpha\;,
\end{split}
\end{equation}
which when plugged back into the curve, with the vanishing condition imposed, gives
\begin{equation}
\begin{split}
\left(\frac{1}{A} - 1 + \alpha - \alpha\right)^2 = 1 
\Rightarrow A = \frac{1}{2}\;.
\end{split}
\end{equation}

\begin{figure}[t]
\centering
\includegraphics[trim = 0mm 0mm 0mm 0mm, clip, scale=0.6]{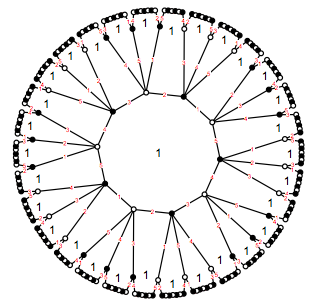}
\caption{\sf {\footnotesize Dimer for the $\mathbb{C}^5$ theory. The ``1" indicates the fact that there is only one $U(N)$ gauge group in this theory, represented by the $5 \times 2 = 10-$sided polygon. The fundamental region encloses 5 black nodes and 5 white nodes, as expected. Figure taken from~\cite{Cremonesi:2013aba}}}
\label{C5}
\end{figure}

The fact that $\alpha$ has not contributed to this calculation tells us that we are free to set it to any value, so choosing $\alpha = 0$ for simplicity, we reveal the final form for the Belyi pair as 
\begin{equation} \label{eq:B5}
y^2 = x^5 +1, \;\;\;\;\;\; \beta\left(y\right) = \frac{y+1}{2}\;.
\end{equation} 
On verification of the ramification structure, we see that indeed it corresponds to $\begin{Bmatrix}
5 \\ 5 \\ 5
\end{Bmatrix}$. This has now been identified as the so-called $\mathbb{C}^5$ theory~\cite{Cremonesi:2013aba}, for which the {\it dessin} is shown above. What the above has shown us is that with some extremely simplified assumptions, and a single-variable ansatz, it was possible to construct the simplest Belyi pair arising in the case of a genus 2 surface. On comparing Equations~(\ref{eq:BC}) and~(\ref{eq:B5}), we see that there is a remarkable similarity between the two. The Belyi map itself is identical, whereas the curve has only been extended from a degree 3 elliptic to a degree 5 hyperelliptic curve. This is no coincidence, as we shall see in Section 6.

%%%%%%%%%%%%%%%%%%%%%%%%%%%%%%%%%%%%%%%%%%%%%%%%%%%%%%
\subsection{Number Field Parameterisations}
Looking back at the initial form of an elliptic curve indicated in Equation~(\ref{eq:ecurve}), we have thus far only considered algebraic descriptions of tori defined over the complex plane $\mathbb{C}$. This is no doubt the most natural field of numbers we could define the curve over, yet relaxing this constraint can actually lead to a rich world of alternative parameterisations of the same. Indeed, rather than limiting ourselves to curves defined over $\mathbb{C}$, we will now look at how the same Belyi pair can be defined over a specific number field~\cite{Elkin}. To illustrate this point, we will once again consider the example of $\begin{Bmatrix}
5 \\ 5 \\ 5
\end{Bmatrix}$ above. We denote the number field as $\mathbb{F}$, defined such that
\begin{equation} \label{eq:NF}
\begin{split}
\xi \in \mathbb{F}: \xi^2 + 2 &= 0  \\
y^2 = 2x^5 - 2, \;\;\;\;\;\;\; \beta&\left(x, y\right) = \frac{\left(2 - \xi y\right)}{x^5}\;.
\end{split}
\end{equation}
$\xi$ here is a complex variable that satisfies the field $\mathbb{F}$. Of course, the solution to this equation is simply $\xi = \pm i\sqrt{2}$. By definition, the Belyi pair (the hyperelliptic curve and the map) are both defined strictly over the number field $\mathbb{F}$. Let us verify that the ramification index of 1 is indeed 5. The cases for 0 and $\infty$ can be calculated analogously. We first begin by substituting the form of the hyperelliptic curve into the map, and rewrite it as \begin{equation}
\beta\left(y\right) = \frac{2\left(2 - \xi y\right)}{y^2 + 2}\;.
\end{equation}
This has allowed us to reduce the map to a function of $y$ only. To locate the preimage of 1, as before, we need to simultaneously solve $\beta\left(y\right) = 1$ and $y^2 = 2x^5 - 2$. This is easily verified to correspond to the point $\left(0, -\xi\right)$ (where we have imposed the condition of the number field, $\xi^2 + 2 = 0$). Our next step is to verify the order of the map at deviations around good local coordinates. As before, we substitute $x = 0 + \delta x$, $y = -\xi + \delta y$ into the hyperelliptic curve to give us
\begin{equation} \label{eq:quad}
\begin{split}
\left(-\xi + \delta y\right)^2 = 2\left(0 + \delta x\right)^5 - 2 \\
\Rightarrow  \left(\delta y\right)^2 - 2 \xi \left(\delta y\right) - 2 \left(\delta x\right)^5 + \xi^2 + 2 = 0\;.
\end{split}
\end{equation}
This is now a quadratic equation in $\delta y$. This time unfortunately, we are not at liberty to choose either the positive or negative solution. We will consider each case separately. Using the negative solution first, we have $\delta y = - \sqrt{2\left(-1 + \left(\delta x\right)^5\right)} + \xi$. To avoid some clutter, let $\delta x = \epsilon$, some small parameter. Now, Taylor expanding to a tenth order in $\epsilon$, we get
\begin{equation}
\delta y = \xi -i \sqrt{2} + \frac{i \epsilon^5}{\sqrt{2}} + \frac{i \epsilon^{10}}{4 \sqrt{2}} + \mathcal{O}\left(\epsilon^{15}\right)\;.
\end{equation}
Now, substituting $x = 0 + \epsilon$, $y = -\xi  + \xi -i \sqrt{2} + \frac{i \epsilon^5}{\sqrt{2}} + \frac{i \epsilon^{10}}{4 \sqrt{2}}$ into the form of the map in Equation~(\ref{eq:NF}), we get
\begin{equation}
\beta = \frac{2 - \left(-i \sqrt{2} + \frac{i \epsilon^5}{\sqrt{2}} + \frac{i \epsilon^{10}}{4 \sqrt{2}}\right)\xi}{\epsilon^5}\;.
\end{equation}
Finally, to simplify the above, we use the {\it positive} solution from the number field $\mathbb{F}$ that $\xi = +i \sqrt{2}$, and get that 
\begin{equation}
\beta = 1 + \frac{\epsilon^5}{4}\;.
\end{equation}
So indeed we see that the map around the ramified point 1 is of order 5. If we had instead used the {\it positive} solution for $\delta y$ in Equation~(\ref{eq:quad}), then we would have to use the {\it negative} solution from $\mathbb{F}$ that $\xi = -i \sqrt{2}$. This may seem alarming at first, but in fact, is a general pattern in this subject -- that conjugate values can lead to different {\it dessins} (i.e., different ramification structures).   Calculations for the ramification indices for 0 and $\infty$ follow in an analogous manner. The calculation illustrated above has been slightly more involved than that for $\mathbb{C}^3$, or indeed than what was done in Section 4.1. In the case of $\begin{Bmatrix}
5 \\ 5 \\ 5
\end{Bmatrix}$, which we have been considering, the standard definition of the Belyi pair over $\mathbb{C}$ is far simpler and a more natural consideration. However, what the number field approach does is that it opens up a huge new set of parameterisations for a Belyi pair. It should theoretically be possible to reformulate any Belyi pair in this manner, and the hope is that this formalism can be used to construct Belyi pairs where the canonical computational rules are difficult when defined over $\mathbb{C}$.

%%%%%%%%%%%%%%%%%%%%%%%%%%%%%%%%%%%%%%%%%%%%%%%%%%%%%%%%%%%%%
\subsection{Igusa Invariants}
In Section~\ref{jinvsec}, we briefly spoke about $j-$invariants, and how they can help distinguish sets of elliptic curves. While Equation~(\ref{eq:jinv}) does not carry through to genus 2 elliptic curves, there is an analogue of the same, known as Igusa invariants~\cite{Igusa}, which are further studied algorithmically in \cite{Mestre,Streng2010,Lauter2000,Lauter2006}. Note that, however, the choice of these invariants, unlike the j-invariant, is not canonical.
Briefly, given the genus 2 hyperelleiptic curve, with the right hand side being a sextic:
\begin{equation}
y^2 = u_0 \prod\limits_{i=1}^6 (x - x_i) \ ,
\end{equation}
where $u_0$ is a complex coefficient are $x_i$ are the 6 roots of the sextic, Igusa defined define the four invariants
\begin{align}
\nn
A' &= u_0^2 \sum\limits_{15 \mbox{perms}} (12)^2(34)^2(56)^2 \ ;
\\
\nn
B' &= u_0^4 \sum\limits_{10 \mbox{perms}} (12)^2 (23)^2 (31)^2(45)^2 (56)^2 (64)^2 \ ;
\\
\nn
C' &= u_0^6 \sum\limits_{60 \mbox{perms}} (12)^2 (23)^2 (31)^2(45)^2 (56)^2 (64)^2 (14)^2 (25)^2 (36)^2 \ ;
\\
\nn
D' &= u_0^{10} \prod\limits_{i<j} (ij)^2 \ .
\end{align}
Here, $(ij)$ is shorthand for $(x_i - x_j)$ and the sums are over the various permutation possible of combining the 6 six roots in pairs as indicated. Indeed, $A'$ are summed over the 15 cross-channels on 6 elements and $D'$ is the discriminant of the sextic. We note that these 4 invariants are of degree $m=2,4,6,8$ respectively and can be compactly written as $f_{m} := u_0^m \sum (x_i - x_j)(x_k - x_{\ell}) \ldots$ in which every $x_i$ appears exactly $m$ times and that $f_m$ is symmetric in all six $x_i$.

Computationally, it is often expedient to define the following vector of invariants, the notation of which is suggestive of the generalisation of the j-invariant,
\begin{equation}
\mbox{igusa} = \left[J_2, J_4, J_6, J_8, J_{10}\right] \;,
\end{equation}
where the $J_i$'s are defined \cite{Mestre} (sometimes called the Igusa-Clebsch invariants) as
\begin{align}
\nn
J_2 &= 2^{-3} A' \ ;
\\ \nn
J_4 &= 2^{-5} 3^{-1} (4 J_2^2 - B') \ ;
\\ \nn
J_6 &= 2^{-6} 3^{-2} (8 J_2^3 - 160 J_2 J_4 - C') \ ;
\\ \nn
J_8 &= 2^{-2} (J_2 J_6 - J_4^2) \ ;
\\ \nn
J_{10} &= 2^{-12} D' \ .
\end{align}
Crucially, a vector of Igusa invariants completely specifies a hyperelliptic curve, up to isomorphism.  We make use of the computational algebra system {\tt MAGMA} to generate the vector of Igusa invariants for the two hyperelliptic curves generating {\small $\begin{Bmatrix}
5 \\ 5 \\ 5
\end{Bmatrix}$}. We get for $y^2 = x^5 + 1$,
\begin{equation}
\mbox{igusa} = \left[0,\, 0,\, 0,\, 0,\, 800000\right] \;,
\end{equation}
whereas for the number field version $y^2 = 2x^5 - 2$,
\begin{equation}
\mbox{igusa} = \left[0,\, 0,\, 0,\, 0,\, 819200000\right] \;.
\end{equation}
The fact that the vector of Igusa invariants is different for the two tells us that despite the visual similarity of the two curves, they are distinct, and represent two distinct genus 2 tori. In other words, there is no coordinate transformation that links the two curves. This is a comforting and exciting result; via two different approaches (over the complex plane first, and then over a number field) and using two distinct hyperelliptic curves, we have been able to find two separate Belyi pairs representing the same ramification structure.

%%%%%%%%%%%%%%%%%%%%%%%========================
%%%%%%%%%========================================
\section{Higher Genera Extensions}
\label{s:higher}
\setall
We commented at the end of Section 4.1 on the remarkable similarity between the Belyi pair constructed in the case of $\begin{Bmatrix}
5 \\ 5 \\ 5
\end{Bmatrix}$ (Equation~(\ref{eq:B5})) and that of $\mathbb{C}^3$ (Equation~(\ref{eq:BC})). The map was exactly the same in both cases, and the only difference was that the order of the polynomial in $x$ defining the curve had been changed to match the degree of the map in each case. Let us now see what happens when we consider the functions\begin{equation}
y^2 = x^7 +1, \;\;\;\;\;\; \beta = \frac{y + 1}{2}\;.
\end{equation}If we now go forth and calculate the ramification structure for this (we do not know yet if it is a Belyi pair or not), then it turns out to be $\begin{Bmatrix}
7 \\ 7 \\ 7
\end{Bmatrix}$. If the pattern above is to follow, given that that we have three ramification points, and that the degree of the map is now 7, then using the Riemann-Hurwitz relation from Equation~(\ref{eq:RH}), we see that this indeed is a Belyi pair, corresponding to genus 3. Without any work at all, we have been able to generate the simplest Belyi pair for the next highest genus. Indeed, it turns out that if we change the degree of the map now to 9, 11 or 13 (and the order of the curve accordingly), then we generate the simplest Belyi pairs for genus 4, 5 and 6 respectively. Note that the reason for using only odd orders for the map is because we limit ourselves to cases with only three ramification points. If we consider even orders (ramification indices) then there is no integer solution for $g$ (the genus of the surface) in the Riemann-Hurwitz relation:
\begin{equation} 
\begin{split}
2g - 2 = \left(\mbox{even degree}\right) - 3 \\ 
\Rightarrow g \notin \mathbb{Z}\;.
\end{split}
\end{equation}
Hence, we can summarise our extension algorithm as follows: for a Riemann surface of arbitrary genus $g$, the Belyi pair for $\begin{Bmatrix}
2g + 1 \\ 2g + 1 \\ 2g + 1
\end{Bmatrix}$ is given by 
\begin{equation}
y^2 = x^{2g +1} +1, \;\;\;\;\;\; \beta = \frac{y + 1}{2}\;.
\end{equation}
The reason why this generalisation is so simple is that in every case, the preimages of 0, 1 and $\infty$ are given by $\left(0, -1\right)$, $\left(0, 1\right)$ and $\left(\infty, \infty\right)$ respectively, exactly as they were in the case of $\mathbb{C}^3$. As a result, the verification of the ramification indices, as set out in Equation 9 follows in exactly the same manner, except that the orders of the variations around the good local coordinates vary according to the genus $g$ as $2g + 1$. This therefore, is a very simple ``genus generalisation" algorithm. While it is only applicable to Belyi pairs involving only three ramification points, it will hopefully serve as a building block to the discovery of even more such rules.

%%%%%%%%%%%%%%%%%%%%%%%%%%%%%%%%%%%%%%%%%%%%%%%%%%%%%%%%%
\section{Genus 3}
\label{s:g3}
\setall
The discussion in Section 5 tells us therefore that the simplest genus 3 Belyi pair comes in the form of $\begin{Bmatrix}
7 \\ 7 \\ 7
\end{Bmatrix}$, for which the pair itself is given by \begin{equation}
y^2 = x^7 +1, \;\;\;\;\;\; \beta = \frac{y + 1}{2}\;.
\end{equation} as demonstrated earlier. Given the trend in identification of $\mathbb{C}^3$ and $\mathbb{C}^5$ in genus 1 and 2 respectively, we may name this theory ``$\mathbb{C}^7$".

%%%%%%%%%%%%%%%%%%%%%%%%%%%%%%%%%%%%%%%%%%%%%%%%
\subsection{Shioda Invariants}
Hyperelliptic curves of genus $g = 3$ (i.e., those in which the curve $y^2 = f\left(x\right)$ has a polynomial $f\left(x\right)$ of degree 7 or 8) are classified by a space of 9 Shioda invariants. As in the case of the $j-$invariants for $g = 1$ or Igusa invariants in $g = 2$, it is possible to reconstruct a genus 3 hyperelliptic curve given just the set of Shioda invariants~\cite{Shioda}:
\vspace{5pt}
\begin{thm}The graded ring $\mathcal{S}$ of invariants of binary octavics is generated by 9 elements $ J_2, J_3, \dots, J_{10}$.
\end{thm}
Within this set, the first 6 invariants remain algebraically independent, whereas the last 3 are related to the others by 5 algebraic relations. In a manner akin to the Igusa invariants, we denote the set as $$\mbox{Shioda} = \left[J_2,J_3,\dots,J_9,J_{10}\right].$$ The subscript $i$ in each $J_i$ represents the weight of that invariant.

As with the Igusa invariant, one can explicitly write these out. Unfortunately, the expressions are quite overwhelming. To give an idea, for the genus 3 hyperelliptic curve given as the general octic:
\begin{equation}
y^2 = a_8 x^8 + a_7 x^7 + \ldots + a_0 \ ,
\end{equation}
we have that
\begin{align}
\nn
J_2 &= \frac{1}{140} ( 280 a_0 a_8 - 35 a_1 a_7 + 10 a_2 a_6 - 5 a_3 a_5 + 2 a_4^2 ),
\\
\nn
J_3 &= \frac{1}{137200} ( 11760 a_0 a_4 a_8 - 7350 a_0 a_5 a_7 + 3150 a_0 a_6^2 - 7350 a_1 a_3 a_8 + 2205 a_1 a_4 a_7 - 525 a_1 a_5 a_6 +
\\
& \quad + 3150 a_2^2a_8 - 525 a_2 a_3 a_7 - 330 a_2 a_4 a_6 + 225 a_2 a_5^2 + 225 a
_3^2a_6 - 135 a_3 a_4 a_5 + 36 a_4^3)
\end{align}
for the first two invariants.
The reader is referred to a fuller treatment of the subject, and a description of the generators of the $J_i$ in~\cite{Shioda,Lercier}. If one considers the set of Shioda invariants as representing a point in the projective space of the given hyperelliptic curve (a 7 or 8-dimensional weighted projective space), then it is possible to normalise this point. To this end, we alternatively identify a set of normalised Shioda invariants, which we denote as $$\mbox{Shioda}^N = \left[j_2,j_3,\dots,j_9,j_{10}\right].$$Given this, the isomorphism condition in genus 3 becomes as follows: two hyperelliptic curves are unique up to isomorphism if they share the same set of {\it normalised} Shioda invariants. Employing the use of {\tt MAGMA}, a standard example would be of the curve $y^2 = x^7 + 1$, for which the set of normalised invariants is: $$\mbox{Shioda}^N = \left[ 0, 0, 0, 0, 0, 1, 0, 0, 0 \right].$$

%%%%%%%%%%%%%%%%%%%%%%==============================
%%%%%%%%%%%%%%%%%%%%%%==============================
\section{Conclusions and Outlook}
\label{s:conc}

A vast class of supersymmetric, four-dimensional gauge theories, by far the largest known to AdS/CFT, is toric in nature by having their moduli space of vacua being non-compact toric Calabi-Yau manifolds.
It is well-established by now that they can be completely encoded by a bipartite graph on a torus known as a dimer model, or equivalently a brane tiling on the doubly-periodic plane.  
It is further known that such combinatorial objects can, number theoretically, be recast in the form of a Belyi pair -- the combination of a rational map and an elliptic curve, an algebraic description of the torus to which the former maps. 

In this work, we have initiated the study of these Belyi pairs for Riemann surfaces of arbitrary genera. The case of genus 1 Belyi pairs has, as mentioned, been extensively studied in literature, with the canonical example of the $\mathbb{C}^3$ theory denoting the well-known case of $\mathcal{N} = 4$ Super-Yang-Mills Theory. Through the definition of the $j$-invariants of an elliptic curve, it is also possible to find alternative algebraic descriptions of the same geometric surface (on which the theory is embedded) simply by means of a coordinate transformation. In addition, we have also seen how the combination of different maps can help to generate new Belyi pairs, such as the geometrically-rich $PdP_4$ theory.

Extending beyond the genus 1 case, we also consider the construction of arbitrary Belyi pairs and their associated gauge theories, giving explicit examples for genus 0, 2 and 3 Riemann surfaces.
Using a combination of available databases and further algorithmic studies, especially with the help of working over finite number fields, we give a catalogue, graded by degree, of these Belyi pairs.
Those theories in which the identification of the Belyi pair currently evades computation can hopefully be described more readily through this alternative approach of working over more intricate fields. We also establish a set of rules allowing for the generation of the simplest Belyi pairs in arbitrary genera and thus give a family of Belyi pairs persisting through genera.

The genus 0 case is perhaps most studied in the mathematics literature and has recently been found to have implications to an interplay between gauge theories and the modular group.
The genus 2 and 3 cases have their own analogues of the j-invariants, the so-called Shioda and Igusa invariants, and we have exploited their properties in identifying the gauge theories.

Physically, dimers on Riemann surfaces of general genera arise in several contexts, ranging from the untwisting procedure in zig-zag paths for toric gauge theories to the recent flurry of activity on encoding scattering amplitudes in ${\cal N} = 4$ using the combinatorics of amplituhedra.
We hope our catalogue of explicit Belyi pairs for these bipartite graphs can be of use to these diverse communities.

\section*{Acknowledgments}
SB is supported by STFC through grant ST/K501979/1, and acknowledges an undergraduate research bursary jointly awarded in 2012 by the Institute of Physics and the Nuffield Foundation, as well as St. Catherine's College, Oxford, and the Institute for Computational Cosmology, Durham University, for their lasting support.
JG is supported by STFC through grant ST/K501906/1, and thanks Trinity College, Cambridge, for a studentship in Mathematics, and Somerville College, Oxford, for their lasting support. 
YHH would like to thank the Science and Technology Facilities Council, UK, for an
Advanced Fellowship and for STFC grant ST/J00037X/1, the Chinese Ministry of
Education, for a Chang-Jiang Chair Professorship at NanKai University, the city of
Tian-Jin for a Qian-Ren Scholarship, the US NSF for grant CCF-1048082, as well as
City University, London, the Department of Theoretical Physics and Merton College,
Oxford, for their enduring support.

~\\
~\\

\bibliographystyle{ieeetr}
\begin{singlespace}
%\begin{thebibliography}{99}
\bibliography{library}
\end{singlespace}
%\include{bibtemp}
%\end{thebibliography}

\newpage

%%%%%%%%%%%%%%%%%%%%%%==============================
%%%%%%%%%%%%%%%%%%%%%%==============================
\appendix

\section{Catalogue of Genus 0 Belyi Maps}\label{a:g0}
In this Appendix, we make a classification of Belyi pairs arising in genus 0, up to degree = 7, as described in Section \ref{s:g0}. Note that since some of the ramification structures have indices equal to 1, not all of them necessarily translate to physically relevant gauge theories. Nonetheless, such a catalogue can prove useful to the mathematical community.

\subsection{Degree 3}
\begin{table}[!ht]
\begin{tabular}{|@{}lc|} 
\hline
\hline
Structure & Map \\ 
\hline 
$\begin{Bmatrix}
3 \\ 3 \\ 1,1,1
\end{Bmatrix}$ & $\frac{(-3i+\sqrt{3})x^3}{9(i+\sqrt{3}-2ix)(x-1)}$ \\ \hline \\
$\begin{Bmatrix}
2,1 \\ 2,1 \\ 3
\end{Bmatrix}$ & $\frac{-27}{4}\left(x - 1\right)x^2$ \\
\hline
\end{tabular}
\end{table}

\newpage
\subsection{Degree 4}
\begin{table}[!ht]
\begin{tabular}{|@{}lc|} 
\hline
\hline
Structure & Map \\ 
\hline 
$\begin{Bmatrix}
4 \\ 4 \\ 1,1,1,1
\end{Bmatrix}$ & $\frac{x^4}{4 - \left(8 - 8i\right)x - 12i x^2 + \left(4 + 4i\right)x^3}$ \\ \hline \\
$\begin{Bmatrix}
3,1 \\ 3,1 \\ 3,1
\end{Bmatrix}$ & $\frac{-4\left(x - 1\right)x^3}{x - \frac{1}{4}}$ \\ \hline \\
$\begin{Bmatrix}
2,2 \\ 2,2 \\ 2,2
\end{Bmatrix}$ & $\frac{-\left(x - 1\right)^2 x^2}{\left(x - \frac{1}{2}\right)^2}$ \\ \hline \\
$\begin{Bmatrix}
2,2 \\ 2,2 \\ 3,1
\end{Bmatrix}$ & No map due to Frobenius formula \\ \hline \\
$\begin{Bmatrix}
3,1 \\ 3,1 \\ 2,2
\end{Bmatrix}$ & $\frac{-64\left(3+2\sqrt{3}\right)\left(-1+x\right)x^3}{9\left(-2+\sqrt{3}+4x\right)^2}$ \\
\hline
\end{tabular}
\end{table}

\newpage

\subsection{Degree 5}
\begin{table}[!ht]
\begin{tabular}{|@{}l c|} 
\hline
\hline
Structure & Map \\ 
\hline 
$\begin{Bmatrix}
4,1 \\ 4,1 \\ 3,1,1
\end{Bmatrix}$ & $\frac{-3125\left(-1+x\right)x^4}{8\left(6+25x\left(-2+5x\right)\right)}$ \\ \hline \\
$\begin{Bmatrix}
4,1 \\ 4,1 \\ 2,2,1
\end{Bmatrix}$ & $\frac{-\left(19-\frac{41i}{2}\right)\left(-1+x\right)x^4}{\left(\left(-4+4i\right)+\left(2-14i\right)x+\left(2+11i\right)x^2\right)^2}$ \\  \hline \\
$\begin{Bmatrix}
3,2 \\ 3,2 \\ 3,1,1
\end{Bmatrix}$ & $\frac{-3125\left(-1+x\right)^2x^3}{-64+100x\left(-4+5x\right)}$ \\ \hline \\
$\begin{Bmatrix}
3,2 \\ 3,2 \\ 2,2,1
\end{Bmatrix}$ & $\frac{3125\left(-1+x\right)^2x^3}{\left(108+25x\left(-9+5x\right)\right)^2}$ \\ \hline \\
$\begin{Bmatrix}
3,1,1 \\ 3,1,1 \\ 5
\end{Bmatrix}$ & $\frac{-25\left(-75i+61\sqrt{15}\right) + \left(7i+\sqrt{15}-8i x\right)\left(-1+x\right)x^3}{2304}$ \\ \hline \\
$\begin{Bmatrix}
2,2,1 \\ 2,2,1 \\ 5
\end{Bmatrix}$ & $\frac{25}{8}\sqrt{5}\left(-1+x\right)^2x^2\left(-1+\sqrt{5}+2x\right)$ \\
\hline
\end{tabular}
\end{table}

\newpage
%\begin{landscape}
\subsection{Degree 6}
\begin{table}[!ht]
\begin{tabular}{|@{}l c|} 
\hline
\hline
Structure & Map \\ 
\hline 
$\begin{Bmatrix}
6 \\ 6 \\ 1,1,1,1,1,1
\end{Bmatrix}$ & $\frac{x^6}{\left(1+\left(-1+x\right)x\right) \left(1+3\left(-1+x\right)x\right)\left(-1+2x\right)}$\\ \hline \\
$\begin{Bmatrix}
5,1 \\ 5,1 \\ 3,1,1,1
\end{Bmatrix}$ & $\frac{-729\left(-1+x\right)x^5}{\left(-1+6x\right)\left(2+15x\left(-1+3x\right)\right)}$\\ \hline \\
$\begin{Bmatrix}
2,4 \\ 2,4 \\ 3,1,1,1
\end{Bmatrix}$ & $\frac{\left(-1+x\right)^4\left(2+x\right)^2}{8x\left(-3+x^2\right)}$ \\ \hline \\
$\begin{Bmatrix}
3,3 \\ 3,3 \\ 3,1,1,1
\end{Bmatrix}$ & No map exists due to Frobenius formula \\ \hline \\
$\begin{Bmatrix}
5,1 \\ 5,1 \\ 2,2,1,1
\end{Bmatrix}$ & $\frac{i x^5\left(-2+\left(2+i\right)x\right)}{\left(-1+2x\right)\left(i+\left(1+2i\right)\left(-1+x\right)x\right)^2}$ \\ 
\hline
\end{tabular}
\end{table}

\begin{landscape}
\begin{table}[!h]
\begin{tabular}{|@{}l c|} 
\hline
\hline
Structure & Map \\ 
\hline 
$\begin{Bmatrix}
2,4 \\ 2,4 \\ 2,2,1,1
\end{Bmatrix}$ & $\frac{-\left(2+\sqrt{3}\right) \sqrt{-3+2 \sqrt{3}} x^4}{\left(1+\sqrt{-3+2 \sqrt{3}}-2 x\right)^2} \times\frac{\left(-3+\sqrt{-9+6 \sqrt{3}}+2 x\right)^2}{\left(\sqrt{3}+\sqrt{3+2 \sqrt{3}}-2 x \left(-3+\sqrt{3+2 \sqrt{3}}+3 x\right)\right)
}$ \\ \hline \\
$\begin{Bmatrix}
3,3 \\ 3,3 \\ 2,2,1,1
\end{Bmatrix}$ & $\frac{\left(-2+\left(-1\right)^{1/3}\right)\left(-1+x^2\right)^3}{9x^2\left(-1+\left(-1\right)^{1/3}+x^2\right)}$\\ \hline \\
$\begin{Bmatrix}
4,1,1 \\ 4,1,1 \\ 5,1
\end{Bmatrix}$ & $\frac{x^4\left(5+2\left(-3+x\right)x\right)}{-1+2x}$ \\  \hline \\
$\begin{Bmatrix}
4,1,1 \\ 4,1,1 \\ 2,4
\end{Bmatrix}$ &  $\frac{10 x^3 \left(20 \left(3+2 \sqrt{6}\right) \left(3-i \sqrt{-9+4 \sqrt{6}}\right)-
15 \left(27+12 \sqrt{6}-i \sqrt{1791+744 \sqrt{6}}\right) x +2 i \sqrt{15 \left(921+376 \sqrt{6}\right)} x^3+
6 x^2 \left(27+12 \sqrt{6}+-i \sqrt{15 \left(921+376 \sqrt{6}\right)}\right)\right)}{3 \left(-15 i+\sqrt{5 \left(3+8 \sqrt{6}\right)}+30 i x\right)^2}$ \\ \hline \\
$\begin{Bmatrix}
4,1,1 \\ 4,1,1 \\ 3,3
\end{Bmatrix}$ & $\frac{\left(-1+x\right)^4\left(1+x\left(4+x\right)\right)}{32x^3}$ \\ \hline \\
$\begin{Bmatrix}
3,2,1 \\ 3,2,1 \\ 5,1
\end{Bmatrix}$ & $\frac{3125\left(x-1\right)^2x^3\left(4+5x\right)}{432\left(-1+5x\right)}$ \\ 
\hline
\end{tabular}
\end{table}

\begin{table}[h]
\begin{tabular}{|@{}l c|} 
\hline
\hline
Structure & Map \\ 
\hline 
$\begin{Bmatrix}
3,2,1 \\ 3,2,1 \\ 2,4
\end{Bmatrix}$ & Not found due to computational limitations \\ \hline \\
$\begin{Bmatrix}
3,2,1 \\ 3,2,1 \\ 3,3
\end{Bmatrix}$ & Not found due to computational limitations \\ \hline \\
$\begin{Bmatrix}
2,2,2 \\ 2,2,2 \\ 5,1
\end{Bmatrix}$ & No map exists due to Frobenius formula \\ \hline \\
$\begin{Bmatrix}
2,2,2 \\ 2,2,2  \\ 2,4
\end{Bmatrix}$ & No map exists due to Frobenius formula \\ \hline \\
$\begin{Bmatrix}
2,2,2 \\ 2,2,2 \\ 3,3
\end{Bmatrix}$ & $\frac{x^2 \left(3+\left(-3+x\right) x\right)^2}{4 \left(-1+x\right)^3}$ \\  
\hline
\end{tabular}
\end{table}
\end{landscape}
 
 \newpage
 
\subsection{Degree 7}
\begin{table}[!ht]
\begin{tabular}{|c|p{12cm}|} 
\hline
\hline
Structure & \qquad \qquad \qquad \qquad \qquad \qquad \qquad \qquad \qquad Map \\ 
\hline 
$\begin{Bmatrix}
7\\ 7 \\ 1,1,1,1,1,1,1
\end{Bmatrix}$ & $\frac{x^7}{1+7\left(-1+x\right)x\left(1+\left(-1+x\right)x\right)^2}$ \\ \hline \\
$\begin{Bmatrix}
6,1 \\ 6,1 \\ 3,1,1,1,1
\end{Bmatrix}$ & $\frac{\left(7-2 x\right) x^6}{5+7 \left(-1+x\right) x \left(4+5 \left(-1+x\right) x\right)}$ \\ \hline \\
$\begin{Bmatrix}
5,2\\ 5,2 \\ 3,1,1,1,1
\end{Bmatrix}$ & $-\frac{\left(7-4 x\right)^2 x^5}{-9+7 \left(-1+x\right) x \left(-3+5 \left(-1+x\right) x\right)}$ \\ \hline \\
$\begin{Bmatrix}
4,3 \\ 4,3 \\ 3,1,1,1,1
\end{Bmatrix}$ & $-\frac{x^4 \left(-7+6 x\right)^3}{1+14 \left(-1+x\right) x \left(-1+4 \left(-1+x\right) x\right)}$ \\ \hline \\
$\begin{Bmatrix}
6,1\\ 6,1 \\ 2,2,1,1,1
\end{Bmatrix}$ & Not found due to computational limitations \\ \hline \\
$\begin{Bmatrix}
5,2 \\ 5,2 \\ 2,2,1,1,1
\end{Bmatrix}$ & No map exists due to Frobenius formula \\
\hline
\end{tabular}
\end{table}

\begin{landscape}
\begin{table}[h]
\begin{tabular}{|c|p{15cm}|} 
\hline
\hline
Structure & \qquad \qquad \qquad \qquad \qquad \qquad \qquad \qquad \qquad Map \\ 
\hline 
$\begin{Bmatrix}
2,4,1\\ 2,4,1 \\ 5,1,1
\end{Bmatrix}$ & \begin{eqnarray*}
\left[\left(-\frac{55}{6}+\frac{5 \left(-9+\sqrt{105}\right)^{2/3}}{3 3^{1/3}}+\frac{20}{\left(3 \left(-9+\sqrt{105}\right)\right)^{2/3}}\right) x^4 \right. &+& \\
\left(23-\frac{16 3^{1/3}}{\left(-9+\sqrt{105}\right)^{2/3}}-\frac{4 \left(-9+\sqrt{105}\right)^{2/3}}{3^{1/3}}\right) x^5 &+&  \\ 
\frac{2}{9} \left(-87+\frac{60 3^{1/3}}{\left(-9+\sqrt{105}\right)^{2/3}}+5 \left(3 \left(-9+\sqrt{105}\right)\right)^{2/3}\right) x^6 &+&\\ \frac{4}{63} \left(87-\frac{60 3^{1/3}}{\left(-9+\sqrt{105}\right)^{2/3}}-5 \left(3 \left(-9+\sqrt{105}\right)\right)^{2/3}\right)
\left. x^7\right] \\ \times \frac{1}{\left(\frac{1}{126} \left(3+\frac{24 3^{1/3}}{\left(-9+\sqrt{105}\right)^{2/3}}+2 \left(3 \left(-9+\sqrt{105}\right)\right)^{2/3}\right)-x+x^2\right)} &&
\end{eqnarray*} \\ \hline \\
$\begin{Bmatrix}
2,4,1\\ 2,4,1 \\ 2,4,1
\end{Bmatrix}$ & \begin{eqnarray*} && \left[x^4 \left(-98+7 \left(21-\sqrt{-1+2 \sqrt{2}}+5 \sqrt{-2+4 \sqrt{2}}\right) x \right. \right. - \\ 
&& \left.\left.  7 \left(7-\sqrt{-1+2 \sqrt{2}}+5 \sqrt{-2+4 \sqrt{2}}\right) x^2+2 \sqrt{-1+2 \sqrt{2}} \left(-1+5 \sqrt{2}\right) x^3\right)\right] / \\ && \left(2 \sqrt{-1+2 \sqrt{2}} \left(-1+5 \sqrt{2}\right)+7 \left(7+\sqrt{-1+2 \sqrt{2}}-5 \sqrt{-2+4 \sqrt{2}}\right) x \right. + \\
&& \left. 7 \left(-21-\sqrt{-1+2 \sqrt{2}}+5 \sqrt{-2+4 \sqrt{2}}\right) x^2+98 x^3\right)\end{eqnarray*}\\
\hline
\end{tabular}
\end{table}

\begin{table}[h]
\begin{tabular}{|c|p{15cm}|} 
\hline
\hline
Structure & \qquad \qquad \qquad \qquad \qquad \qquad \qquad \qquad \qquad Map \\ 
\hline 
$\begin{Bmatrix}
5,1,1 \\ 5,1,1 \\ 5,1,1
\end{Bmatrix}$ & $\frac{343\left(7i+11\sqrt{7}\right)\left(-1+x\right)x^5\left(-3+i \sqrt{7}+4x\right)}{64\left(3i+\sqrt{7}-14i x\right)\left(-2+7x\right)}$ \\ \hline \\
$\begin{Bmatrix}
5,1,1 \\ 5,1,1 \\ 2,4,1
\end{Bmatrix}$ & \begin{eqnarray*} && \left[ x^5 \left(7 \left(-21+\sqrt{21 \left(69-8 i \sqrt{6}\right)}\right) + \right. \right. \\ 
&& \left. \left. x \left(49-7 \sqrt{21 \left(69-8 i \sqrt{6}\right)}+2 \sqrt{21 \left(69-8 i \sqrt{6}\right)} x\right)\right) \right. / \\ 
&& \left(2 \left(49+\sqrt{21 \left(69-8 i \sqrt{6}\right)}\right)-7 x \left(63+\sqrt{21 \left(69-8 i \sqrt{6}\right)}- \right. \right.\\ 
&&  \left. \left. \left(105+\sqrt{21 \left(69-8 i \sqrt{6}\right)}\right) x+70 x^2\right)\right) \end{eqnarray*}\\ \hline \\
$\begin{Bmatrix}
5,1,1\\ 5,1,1 \\ 3,2,2
\end{Bmatrix}$ & $\frac{x^5\left(6\left(2+\sqrt{3}\right)+x\left(-7-2\sqrt{3}+2x\right)\right)}{\left(2-\sqrt{3}+x\left(-5+2\sqrt{3}+5x\right)\right)^2}$ \\  \hline \\
$\begin{Bmatrix}
5,1,1 \\ 5,1,1 \\ 3,3,1
\end{Bmatrix}$ & Not found due to computational limitations \\ 
\hline
\end{tabular}
\end{table}

\begin{table}[h]
\begin{tabular}{|c|p{15cm}|} 
\hline
\hline
Structure & \qquad \qquad \qquad \qquad \qquad \qquad \qquad \qquad \qquad Map \\ 
\hline 
$\begin{Bmatrix}
3,2,2\\ 3,2,2 \\ 3,2,2
\end{Bmatrix}$ & $\frac{x^3\left(7-7x+4x^2\right)^2}{\left(4-7x+7x^2\right)^2}$ \\ \hline \\
$\begin{Bmatrix}
3,2,2 \\ 3,2,2 \\ 3,3,1
\end{Bmatrix}$ & 
\begin{eqnarray*} && \left[49 \left(35+10 \sqrt{21}-7 i \sqrt{5 \left(31-4 \sqrt{21}\right)}-40 x\right)^2 \times \right. \\
&& \left. \left(125 i \left(-7+2 \sqrt{21}\right)-7 \sqrt{5 \left(31-4 \sqrt{21}\right)} \left(31+4 \sqrt{21}\right)+.1000 i x\right)^2 x^3 \right]\times \\ &&\frac{1}{\left(50000 \left(-35-i \sqrt{35}+70 x\right) \left(-35+17 i \sqrt{35}+70 x\right)^3\right)} \end{eqnarray*} \\ \hline \\
$\begin{Bmatrix}
3,2,2\\ 3,2,2 \\ 5,1,1
\end{Bmatrix}$ & $\frac{x^3\left(35+12x\left(-7+4x\right)\right)^2}{-1+21\left(-1+x\right)x}$ \\ \hline \\
$\begin{Bmatrix}
3,2,2 \\ 3,2,2 \\ 2,4,1
\end{Bmatrix}$ & $\frac{42 \sqrt{105} x^3 \left(-35 \left(-5+\sqrt{105}\right)+28 \left(-15+\sqrt{105}\right) x+240 x^2\right)^2}{\left(105+13 \sqrt{105}-210x\right) \left(-105+19 \sqrt{105}+210 x\right)^2}$ \\ \hline \\
$\begin{Bmatrix}
3,3,1\\ 3,3,1 \\ 5,1,1
\end{Bmatrix}$ & $-\frac{\left(-1347+145\sqrt{105}\right)\left(35+\sqrt{105}-32x\right)^3 x^3\left(-7+3\sqrt{105}+32x\right)}{1179648\left(5+\sqrt{105}-56\left(-1+x\right)x\right)}$ \\  
\hline
\end{tabular}
\end{table}

\begin{table}[h]
\begin{tabular}{|c|p{15cm}|} 
\hline
\hline
Structure & \qquad \qquad \qquad \qquad \qquad \qquad \qquad \qquad \qquad Map \\ 
\hline 
$\begin{Bmatrix}
4,1,1,1 \\ 4,1,1,1 \\ 7
\end{Bmatrix}$ & $20x^4\left(-7/4+21x/5-7x^2/2+x^3\right)$ \\  \hline \\
$\begin{Bmatrix}
3,2,1,1\\ 3,2,1,1 \\ 7
\end{Bmatrix}$ & \bq && -\frac{1}{45927}\left(7693+2527 2^{2/3} 7^{1/3}+1688 2^{1/3} 7^{2/3}\right)x^3 \\
&& \left(-7+2^{2/3} 7^{1/3}+6 x\right)^2 \left(14-14 2^{2/3} 7^{1/3}+ \right. \\
&& \left. \left(49+7 \cdot 2^{2/3} 7^{1/3}-4 \cdot 2^{1/3} 7^{2/3}\right) x+6 \left(-7+2^{2/3} 7^{1/3}\right) x^2\right) \eq\\ 
\hline
\end{tabular}
\end{table}

\begin{table}[h]
\begin{tabular}{|c|p{15cm}|} 
\hline
\hline
Structure & \qquad \qquad \qquad \qquad \qquad \qquad \qquad \qquad \qquad Map \\ 
\hline 
$\begin{Bmatrix}
2,2,2,1 \\ 2,2,1,1 \\ 7
\end{Bmatrix}$ & 
\begin{eqnarray*} && -\left[\left(289 i \left(-71+39 i \sqrt{3}\right)^{2/3}+7 \cdot14^{1/3} \left(\left(7 \left(-71+39 i \sqrt{3}\right)\right)^{1/3} \left(11 i+23 \sqrt{3}\right) + \right. \right. \right. \\ 
&& \left. \left. 2 \cdot 2^{1/3} \left(-124 i+43 \sqrt{3}\right)\right)\right) x^2 \times \\ 
&& \left(14^{1/3} \left(\left(7 \left(-71+39 i \sqrt{3}\right)\right)^{1/3} \left(-6 i+4 \sqrt{3}\right)+2^{1/3} \left(-3 i+37 \sqrt{3}\right)\right) - \right. \\ 
&& \left. 6 i \left(-71+39 i \sqrt{3}\right)^{2/3} x\right) \times \\ 
&& \left(7 \left(-71+39 i \sqrt{3}\right)^{2/3}+14^{1/3} \left(2^{1/3} \left(37-i \sqrt{3}\right) + \right. \right. \\ 
&& \left. \left. 2 i \left(7 \left(-71+39 i \sqrt{3}\right)\right)^{1/3} \left(2 i+\sqrt{3}\right)\right)+3 \left(-71+39 i \sqrt{3}\right)^{2/3}
x\right)^2 \times \\
&& \left(14^{1/3} \left(2 \cdot 2^{1/3} \left(27-10 i \sqrt{3}\right)+i \left(7 \left(-71+39 i \sqrt{3}\right)\right)^{1/3} \left(9 i+\sqrt{3}\right)\right) + \right. \\ 
&& \left. \left. 6 \left(-71+39 i \sqrt{3}\right)^{2/3} x\right)^2\right] / \left(23328 \left(71 i+39 \sqrt{3}\right)^4\right) \end{eqnarray*} \\
\hline
\end{tabular}
\end{table}
\end{landscape}

%\begin{landscape}
\newpage
\section{Catalogue of Genus 1 Belyi Maps}\label{a:g1}
\begin{table}[!ht]
\begin{tabular}{|c|p{6cm}|p{9cm}|} 
\hline
\hline
Structure & \qquad \qquad \qquad   Elliptic Curve & \qquad \qquad  \qquad \qquad \qquad \qquad Map \\ 
\hline 
$\begin{Bmatrix}
3 \\ 3 \\ 3
\end{Bmatrix}$ & $y^2 = x^3 + 1$ & $\beta \left(x, y\right) = \frac{1}{2} \left(1+y\right)$  \\ \hline \\
$\begin{Bmatrix}
4 \\ 4 \\ 3,1
\end{Bmatrix}$ & \bq && \xi \in \mathbb{F}: \xi^4 - 2 = 0\;, \\
&& y^2 = x^3 + \frac{47}{1944}x + \frac{2359}{314928}\xi^2 \eq & \bq \beta \left(x, y\right) = \frac{1}{\frac{467}{972} + \frac{\xi^2}{9}x - 6x^2 - 4\xi y} \eq \\ \hline \\
$\begin{Bmatrix}
4 \\ 4 \\ 2,2
\end{Bmatrix}$ & $y^2 = x^3 - x$ & $\beta \left(x, y\right) = \frac{ \left(x+1\right)^2}{4x}$ \\ \hline \\
$\begin{Bmatrix}
5 \\ 5 \\ 3,1,1
\end{Bmatrix}$ &\bq && \xi \in \mathbb{F}: \xi^4 - 2\xi^3 - 6\xi^2 - \\&& 8\xi + 16 = 0\;, \\
 && y^2 = x^3 + \frac{25}{324}x + \\ && \frac{1}{839808} \left(-1975\xi^3 6 + \right. \\ 
&&\left. 3950\xi^2 +  19750\xi + 7900\right) \eq &
\bq \beta \left(x, y\right) = && -24300/\left(25  \left(-424+3888 x^2 + \right. \right. \\ 
&& \left. \left. 9 x \left(2+\xi \right) \left(-2+\left(-4+\xi \right) \xi \right)\right) + \right. \\
&& 108 y \left(50 \left(-8+\xi  \left(-2+\left(-2+\xi \right) \xi \right)\right) + \right. \\ 
&& \left. \left. 9 x \left(-32+\xi  \left(-18+\xi  \left(6+\xi \right)\right)\right)\right)\right)\eq \\ 
\hline
\end{tabular}
\end{table}

\begin{landscape}
\begin{table}[h]
\begin{tabular}{|c|p{6cm}|p{14cm}|} 
\hline
\hline
Structure & \qquad \qquad \qquad  Elliptic Curve &  \qquad \qquad  \qquad \qquad \qquad \qquad Map \\ 
\hline 
$\begin{Bmatrix}
5 \\ 5 \\ 2,2,1
\end{Bmatrix}$ & $\xi \in \mathbb{F}: \xi^4 +10 = 0\;, y^2 = x^3 + \frac{5275}{6144}x + \frac{77675}{1769472}\xi^2$ & $\beta\left(x, y\right) = \frac{1}{\frac{17969}{18432} - \frac{\xi^2}{96}x - x^2 -  \left(-\frac{\xi}{24} - \frac{4 \xi^3}{25}x\right) y} $ \\ \hline \\
$\begin{Bmatrix}
4,1 \\ 4,1 \\ 5
\end{Bmatrix}$ & $\xi \in \mathbb{F}: \xi^4 - 45 = 0\;, y^2 = x^3 + 9x + \frac{18}{5}\xi^2$ & \bq && \beta\left(x, y\right) = \frac{\frac{16}{3} \xi^3 x^3 + 1680 \xi x^2 + \frac{9216}{25}\xi^3 x - \frac{29376}{5}\xi}{\left(x^5 - 9\xi^2 x^4 + 1458 x^3 - \frac{13122}{5} \xi^2 x^2 + \frac{531441}{5} x - \frac{4782969}{125} \xi^2\right)y} + \\ \\
&& \frac{4 x^5 + 132 \xi^2 x^4 + 10620 x^3 - 2484 \xi^2 x^2 + \frac{603936}{5}x - \frac{3158352}{125} \xi^2}{\left(x^5 - 9 \xi^2 x^4 + 1458 x^3 - \frac{13122}{5} \xi^2 x^2 + \frac{531441}{5} x - \frac{4782969}{125} \xi^2\right)} \eq\\ \hline \\
$\begin{Bmatrix}
3,2 \\ 3,2 \\ 5
\end{Bmatrix}$ &  
$\xi \in \mathbb{F}: \xi^4 + 40 = 0\;,
y^2 = x^3 + 3x + \frac{37}{80}\xi^2$ & \bq && \beta\left(x, y\right) = \frac{\left(3/2 \xi^3 x^3 + 45/2 \xi x^2 + 153/50 \xi^3 x - 69/5 \xi\right)}{\left(x^5 + 2 \xi^2 x^4 - 64 x^3 - 128/5 \xi^2 x^2 + 1024/5 x + 2048/125 \xi^2\right) y} + \\ \\
&& \frac{\left(4 x^5 - 17/2 \xi^2 x^4 - 10 x^3 - 173/8 \xi^2 x^2 + 679/5 x + 461/500 \xi^2\right)}{\left(x^5 + 2 \xi^2 x^4 - 64 x^3 - 128/5 \xi^2 x^2 + 1024/5 x + 2048/125 \xi^2\right)} \eq\\ \hline  \\ 
$\begin{Bmatrix}
6 \\ 6 \\ 3,1,1,1
\end{Bmatrix}$ & $y^2 = x^3 + 1$ & $\beta \left(x, y\right) = \frac{1}{1- \frac{8+8x^3+x^6}{x^6} - \frac{\left(-8-4x^3\right)}{x^6}y}$ \\
\hline
\end{tabular}
\end{table}

\begin{table}[h]
\begin{tabular}{|c|p{6cm}|p{14cm}|} 
\hline
\hline
Structure & \qquad \qquad \qquad  Elliptic Curve &  \qquad \qquad  \qquad \qquad \qquad \qquad Map \\ 
\hline 
$\begin{Bmatrix}
6 \\ 6 \\ 2,2,1,1
\end{Bmatrix}$ & $y^2 = x^3 + \frac{2}{3}x - \frac{7}{27}$ & $\beta \left(x, y\right) = \frac{1}{1+ \frac{1}{27}\left(1 - 9x + 27x^2-27x^3\right)}$ \\ \hline \\
$\begin{Bmatrix}
3,3 \\ 3,3 \\ 3,3
\end{Bmatrix}$ & $y^2 = x^3 - \frac{15}{16}x + \frac{11}{32}$ & $\beta \left(x, y\right) = \frac{1}{2}\left(1+ \frac{x^2 - x +\frac{7}{16}}{\left(x - \frac{1}{2}\right)^2}y\right)$ \\ \hline \\
$\begin{Bmatrix}
3,3 \\ 3,3 \\ 4,2
\end{Bmatrix}$ & N/A & No Belyi pair exists due to Frobenius formula \\
\hline
\end{tabular}
\end{table}
\end{landscape}

\newpage
\section{Catalogue of Genus 2 Belyi Maps}\label{a:g2}
\begin{table}[!ht]
\begin{tabular}{|@{}l c | c |} 
\hline
\hline
Structure & Elliptic Curve & Map \\ 
\hline 
$\begin{Bmatrix}
5 \\ 5 \\ 5
\end{Bmatrix}$ & $y^2 = x^5 + 1$ & $\beta \left(x, y\right) = \frac{y + 1}{2}$\\ \hline \\
$\begin{Bmatrix}
6 \\ 6 \\ 3,3
\end{Bmatrix}$ & $y^2 = x^6 + 3 x^3 +\frac{1}{4}$ & $\beta \left(x, y\right) = \frac{1}{1+\frac{1}{2}\left(-1-8x^3-4x^6\right) - \left(1+2x^3\right) y}$ \\
\hline
\end{tabular}
\end{table}

\end{document}